\documentclass[prd,aps,a4paper,nofootinbib,twocolumn]{revtex4}  

\newif\ifusesec
\usesectrue  
   
\usepackage{graphicx} 
\usepackage{mathrsfs}
\usepackage{amsmath,amsfonts,amssymb}
\usepackage{multirow}

\newcommand{\beq}{\begin{equation}}
\newcommand{\eeq}{\end{equation}}
\newcommand{\bea}{\begin{eqnarray}}
\newcommand{\eea}{\end{eqnarray}}
\def\pmb#1{\setbox0=\hbox{$#1$}%
  \kern-.025em\copy0\kern-\wd0
  \kern.05em\copy0\kern-\wd0
  \kern-.025em\raise.0433em\box0}
\begin{document}

\title{Scalar waves in a Topological Star  spacetime: self-force and  radiative losses}

\author{Massimo Bianchi$^{1}$, Donato Bini$^{2}$, Giorgio Di Russo$^{1}$}
  \affiliation{
$^1$Dipartimento di Fisica, Universit\`a di Roma \lq\lq Tor Vergata" and Sezione INFN Roma2, Via della
Ricerca Scientifica 1, 00133, Roma, Italy\\
$^2$Istituto per le Applicazioni del Calcolo ``M. Picone'' CNR, I-00185 Rome, Italy\\
}

\date{\today}

\begin{abstract}
We study the radiated energy by a scalar particle moving on a circular orbit (smeared in the extra dimension) in the spacetime of a topological star,  extending a previous study [Phys. Rev. D \textbf{110}, 084077 (2024)]. We discuss motion in the presence of self-force effects too.
\end{abstract}


\maketitle
 
\section{Introduction}

With the growing amount of available data on compact binary mergers
\cite{LigoVirgoKagra} one can hope to discriminate gravitational wave
signals produced by Exotic Compact Objects (ECO's) \cite{Raposo:2018xkf,Barack:2018yly} from those expected
to be emitted by standard Black Holes (BHs) in General Relativity. Among ECO's
a particularly interesting class are the so-called ``fuzzballs,'' that
consist of ensembles of smooth horizonless geometries representing (a
fraction of) the micro-states of the putative BHs \cite{Lunin:2001jy,Bena:2007kg,Skenderis:2008qn}.

Various families of solutions of this kind have been constructed \cite{Bena:2015bea} 
and carefully analysed in terms of their multipolar
structure \cite{Bianchi:2020miz,Bianchi:2020bxa,Bena:2020uup,Bena:2020see}, their (critical) geodesics \cite{Bianchi:2017sds,Bianchi:2020yzr}, 
their quasi-normal modes (QNMs) \cite{Bianchi:2022qph} 
and
tidal deformations \cite{Bianchi:2023sfs,DiRusso:2024hmd}.  

Unfortunately, generic
micro-states may not admit a geometric interpretation and in most of the
cases the dynamics is not integrable \cite{Bianchi:2018kzy,Bena:2019azk}.

This should
not come as a surprise, since these objects should consistently describe
`maximally' chaotic systems like putative BHs \cite{Maldacena:2015waa,Bianchi:2020des}, 
that are
expected to be \lq fast scramblers' \cite{Susskind:2014rva,Cotler:2016fpe}.

In order to test promising ideas for \lq what-should-replace-BHs' an
extremely simple class of models are the so-called Top(ological) Stars (TS),
smooth horizonless solutions of Einstein-Maxwell theory in $D=5$
\cite{Bah:2020pdz}. For specific choices of the parameters, the
solution is capped and can be easily reduced to $D=4$ along a compact
circle. This allows to fully integrate the dynamics of both massive and
massless particles probing the background \cite{Bianchi:2022qph,Heidmann:2023ojf}.  

In \cite{Bianchi:2024vmi} we have studied (massless) scalar waves
produced by a (massive) scalar particle probing the space-time of a
TS. In particular we discussed the analogue of the
relativistic Poynting-Robertson effect and the self-force-driven
evolution of the scalar probe. To this end, we carefully determined the
(critical) geodesics, which turn out to be planar and circular thanks to
spherical symmetry, and solved the relevant wave equation (a Confluent
Heun Equation, CHE) both in the  well established  Mano-Suzuki-Takasugi (MST) approach
\cite{Mano:1996vt} and in the  new approach based on N=2 supersymmetric
gauge theory in a non-commutative Nekrasov-Shatashvili background \cite{Nekrasov:2009rc,Nekrasov:2002qd} or,
equivalently thanks to the Alday-Gaiotto-Tachikawa (AGT) correspondence \cite{Alday:2009aq},
on Liouville Conformal Field Theory \cite{Aminov:2020yma,Aminov:2023jve,Bautista:2023sdf,Bonelli:2022ten,Bonelli:2021uvf,Bianchi:2021xpr,Bianchi:2021mft,Consoli:2022eey,Fucito:2023afe,Cipriani:2024ygw,Bianchi:2023rlt}. 

A remarkable
outcome of our analysis was the perfect identification between the
`renormalized angular momentum' $\nu$ in the MST description and the
fundamental period $a$ of  the `quantum' Seiberg-Witten (qSW) curve in the Supersymmetric Yang-Mills/Conformal Field Theory
(SYM/CFT) approach. In fact $\nu=a-{1\over 2}$ exactly.

The present investigation aims at studying the energy radiated by a
scalar particle circularly orbiting around a TS and discussing
motion in presence of self-force effects.

After a very concise review of the two classes of TSs and their
properties, we study  scalar waves produced   by a
massive scalar source  (actually a small string wound around the compact
$y$ direction, that allows to consistently focus on the $n=0$ Kaluza-Klein (KK) sector) moving on a circular equatorial orbit, 
by using the Green's function method. We then compute the self-force on the
probe along its orbit that allows to derive the modified dynamics
and the radiated energy and angular momentum.

Although we will mostly focus on the $n=0$ KK sector 
and spin zero waves, we perform a rough
comparison with similar results in `standard' BH contexts and comment on
extensions to $n\neq 0$ and $s\neq 0$ waves. We hope to report soon on
gravitational waves that have been partly analyzed with different
vantage points in \cite{Bena:2024hoh,Dima:2024cok}.

We also add two appendices. In Appendix A we discuss the introduction of a tortoise coordinate and study the asymptotic behavior of the solutions. In Appendix B we describe the $r-\phi$ and $r-y$ sections of the TS metrics in order to gain some visualization of the geometry.

\section{Topological Star spacetime}
\label{TSmetric}

A Top(ological)  Star \cite{Bah:2020pdz}  is described by the following metric in $D=5$\footnote{We use mostly positive signature, $-++++$, and coordinates $x^\mu=\{t,r,\theta,\phi,y\}$, $\mu=0,\ldots 4$.}
\bea
\label{metric_top_star}
ds^2 &=& -f_s(r)dt^2+\frac{dr^2}{f_s(r)f_b(r)} \nonumber\\ 
&+& r^2 (d\theta^2+\sin^2\theta d\phi^2) + f_b(r) dy^2\,,
\eea
with 
\beq
f_{s,b}(r)=1-{r_{s,b}\over r}\,.
\eeq
The coordinate $y$ is compact $y \sim y+ 2\pi R_y$.
We denote
\beq
\label{Delta_def}
\Delta(r)=r^2f_s(r)f_b(r)\,,
\eeq
and, driven by the $f_{s,b}(r)$ notation, we define also
\beq
f_{\rm ph}(r)=1-\frac{3  r_s}{2r}\,,\qquad f_{_{_{\rm ISCO}}}(r)=1-\frac{3 r_s}{r}\,,
\eeq
where \lq\lq ph" stands for \lq\lq photon sphere" and \lq\lq ISCO" for \lq\lq Innermost Stable Circular Orbit." 

The geometry \eqref{metric_top_star} represents a magnetically charged solution of 5D Einstein-Maxwell's equations, sourced by the electromagnetic field
\beq
F=P\sin\theta d\theta \wedge d\phi\,,\qquad P^2=\frac{3r_br_s}{2\kappa_5^2}\,, 
\eeq
with $P$ representing a `magnetic' charge\footnote{An `electric' solution corresponding to a string wound around the $y$ direction with $H_3 = Q dt\wedge dy\wedge dr/r^2={}^{*_5} F_2$ is also known.}.
From a 5D perspective the geometry is regular for $r_s$, $r_b$ and $R_y$ satisfying
the condition
\beq
r_s=r_b \left(1-\frac{4r_b^2}{R_y^2}\right)\,,
\eeq
namely
\beq
\label{R_y_def}
R_y=\frac{2r_b^{3/2}}{\sqrt{r_b-r_s}}\,,
\eeq
that implies\footnote{At fixed mass $M_{\rm TS}$ the minimal radius is $R_y = 4\sqrt{2}M_{\rm TS}$ reached for $r_b=2r_s$.} that $r_s \le  r_b$ and $R_y \ge  2r_b$.
Solutions without thermodynamical (or Gregory-Laflamme) instabilities require $r_s < r_b < 2r_s$ and in general belong to two different classes depending on whether $r_b \lessgtr 3r_s/2$ (see {\it e.g.} \cite{Bianchi:2023sfs}).

Type I Top Stars with $r_b>3r_s/2$ display a single unstable photon ring at $r=r_b$. 

Type II Top Stars with $r_b<3r_s/2$ display a stable photon ring at $r=r_b$ and an unstable photon ring at $r=3r_s/2$. For massive probes with mass $\mu$,  angular momentum $K$ and vanishing component of the momentum along the $y$ direction, $p_y=0$, the three critical radii are 
\beq
\label{rcritmassive}
r_{c,0} = r_b \quad , \quad r_{c,\pm} = {K^2 \pm K \sqrt{K^2- 3\mu^2 r_s^2} \over \mu^2 r_s}\,. 
\eeq
For $\mu\ll 1$ and denoting $\hat K=\frac{K}{\mu r_s}$ (dimensionless) \footnote{Correcting a typo in the second equation in (3.24) of \cite{Bianchi:2023sfs}.} 
\bea
\label{rcritmassapprox}
r_{c,+} &=&  2\hat K^2 r_s - {3\over 2} r_s    -\frac98 \frac{r_s}{\hat K^2 }+O(\mu^4) \,,\nonumber\\
r_{c,-} &=& \phantom{2\hat K^2 r_s} +{3\over 2} r_s + \frac98 \frac{r_s}{\hat K^2 } +O(\mu^4)\,.
\eea
The ISCO corresponds to $r_{c,+} =r_{c,-} =3r_s$.

After dimensional reduction to $D=4$, the solution exposes a naked singularity and has a mass
\beq
\label{mass_TS}
4 G_4 M_{\rm TS} = 2 r_s + r_b\,,
\eeq
with 
\beq
8\pi G_4 =\kappa_4^2= {\kappa_5^2\over 2\pi R_y}=\frac{8\pi G_5}{2\pi R_y}\,. 
\eeq

Note in passing that the TS metric \eqref{metric_top_star} admits a Killing-Yano skew-symmetric tensor (exactly as the Schwarzschild Killing-Yano tensor)
\beq
f_{\rm KY}=r^3\sin\theta d\theta \wedge d\phi\,,  
\eeq
with the property $f_{\rm KY}{}_{\alpha (\beta;\gamma)}=0$, such that $f_{\rm KY}\propto P$, i.e., 
\beq
F=P\frac{f_{\rm KY}}{r^3}\,.
\eeq 

Hereafter we will   set $G_4=1=c$.
For $r_b=0$, and thus $r_s=2GM_{\rm TS}$, the resulting singular solution is a Schwarzschild BH times a circle.
We will  resort to this limit when useful.
Finally, it will be convenient to scale $r_b$ by $r_s$, and consequently to define a dimensionless parameter, $\alpha >1$, such that
\beq
r_b=\alpha r_s\,.
\eeq

\section{Scalar waves}

In the TS spacetime \eqref{metric_top_star} let us consider the scalar wave equation for a massless scalar field $\psi$ in absence of couplings with the background but with a source $\rho$ (scalar charge density) due to a massive particle moving on the background along a curve with parametric equations $x^\mu=x_p^\mu(\tau)$
\beq
\label{eq_fund}
\Box \psi=-4\pi \rho\,,
\eeq
where
\beq
\Box \psi=\frac{1}{\sqrt{-g}}\partial_\mu (\sqrt{-g}g^{\mu\nu}\partial_\nu  \psi)\,,
\eeq
and 
\beq
\label{rho_gen_def}
\rho=q\int \frac{d\tau}{\sqrt{-g}} \delta^{(5)}(x-x_p(\tau))\,,
\eeq
with $\sqrt{-g}=r^2\sin\theta$ (and $q$ in 5D with different dimensions of $q$ in 4D, an important feature when making comparisons). The source of the field is then nonzero (and Dirac-delta singular) along the source world line.

\subsection{Source}

As in Ref.~\cite{Bianchi:2024vmi}, we assume the source moving on a timelike circular equatorial geodesic, with parametric equations
\bea
t&=&t_p(\tau)=\Gamma \tau\,,\quad r=r_p(\tau)=r_0\,,\nonumber\\
\theta&=&\theta_p(\tau)=\frac{\pi}{2}\,,\quad\phi=\phi_p(\tau)=\Gamma \Omega \tau\,,\quad y_p(\tau)=0\,,\qquad
\eea
with $r_0>{\rm max}\{r_b,\frac32 r_s\}$ and 
\beq
\label{Omega_circ}
\Omega=\sqrt{\frac{r_s}{2r_0^3}} \,,\qquad \Gamma =\frac{1}{\sqrt{1-\frac{3r_s}{2r_0}}}\,,
\eeq
and four velocity 
\beq
U^\mu =\frac{dx^\mu_p(\tau)}{d\tau}=\Gamma (\delta^\mu_t+\Omega \delta^\mu_\phi)\,.
\eeq
 The associated  energy and angular momentum per unit mass, $\hat E$ and $\hat L$, are given by
\beq
\label{E_circ}
\hat E=\frac{1-\frac{r_s}{r_0}}{\sqrt{1-\frac{3r_s}{2r_0}}}\,,\qquad \frac{\hat L}{M}=\frac{1}{\sqrt{\frac{r_s}{2r_0}\left(1-\frac{3r_s}{2r_0}\right)}}\,,
\eeq
with $\hat K=\frac{\hat L}{r_s}=\frac{\hat L}{2M}$ in the equatorial plane, see Eq. \eqref{rcritmassive}. 

The density $\rho$, Eq. \eqref{rho_gen_def}, then writes as
\bea
\rho&=& q\int \frac{d\tau}{r^2\sin \theta}\delta^{(5)}(x-x_p(\tau)) \,,
\eea
where 
\bea
\label{delta5}
\delta^{(5)}(x-x_p(\tau))&=& \delta(t-\Gamma \tau)\delta(r-r_0)\delta(\theta-\frac{\pi}{2})\times \nonumber\\
&&\times \delta(\phi-\Omega t)\delta(y)\,.
\eea
The integral over $\tau$ is immediate and gives
\bea
\label{rho_def}
\rho&=& \frac{q} {r_0^2\Gamma}\delta(r-r_0)\delta(\theta-\frac{\pi}{2}) \delta(\phi-\Omega t)\delta(y)\nonumber\\
&=& \frac{q} {r_0^2\Gamma} \sum_{lmn}\delta(r-r_0) Y_{lm}(\theta,\phi)e^{-im\Omega t}Y_{lm}^*(\frac{\pi}{2},0)\frac{e^{i n \frac{y}{R_y}}}{2\pi R_y}\nonumber\\
&=& \frac{q} {r_0^2\Gamma} \sum_{lmn}\delta(r-r_0) Y_{lm}(\theta,\phi)Y_{lm}^*(\frac{\pi}{2},0)\times \nonumber\\
&\times & \int \frac{d\omega}{2\pi}e^{-i\omega t}\delta(\omega-m\Omega)\frac{e^{i n \frac{y}{R_y}}}{2\pi R_y}
\eea
where we have used a Fourier series representation  for $\delta(y)$ (instead of the integral representation) since $y$ is a periodic variable
\beq
\delta(y)=\sum_{n=-\infty}^\infty \frac{e^{i n \frac{y}{R_y}}}{2\pi R_y}\,,
\eeq
besides the well known identity 
\bea
\delta(\cos\theta)\delta(\phi-\phi_0)&=&\delta(\theta-\frac{\pi}{2})\delta(\phi-\phi_0)\nonumber\\
&=&\sum_{lm} Y_{lm}(\theta,\phi)Y_{lm}^*(\frac{\pi}{2},\phi_0)\,.
\eea
Consequently one can write the scalar charge density $\rho$ in the form
\bea
\label{rho_t2}
\rho(x) 
&=& \sum_{lmn}{\mathcal S}_{lmn}(t,r)e^{i n \frac{y}{R_y}} Y_{lm}(\theta, \phi)\nonumber\\
&=&  \sum_{lmn}\int_{\omega}  e^{-i\omega t}\widehat {\mathcal S}_{lmn}(\omega,r)e^{i n \frac{y}{R_y}} Y_{lm}(\theta, \phi)
\,, 
\eea
where
\beq
\int_{\omega}=\int \frac{d\omega}{2\pi}\,.
\eeq
Furthermore, for dimensional reasons and to ease comparison with the Schwarzschild spacetime, it is convenient to define
\beq
q_{_{\rm TS}}=\frac{q}{2\pi R_y}\,,
\eeq 
that can be used as an expansion parameter.
Therefore,
\bea
{\mathcal S}_{lmn}(t,r)&=& \int_{\omega}e^{-i\omega t}\widehat {\mathcal S}_{lmn}(\omega, r)\nonumber\\
&=&  \frac{q_{_{\rm TS}}} {r_0^2\Gamma}\delta(r-r_0) e^{-im\Omega t}Y_{lm}^*(\frac{\pi}{2},0)\,,\nonumber\\
\widehat {\mathcal S}_{lmn}(\omega,r)&=&\int dt   e^{i\omega t}   {\mathcal S}_{lmn}(t,r)\nonumber\\
&=&   \frac{q_{_{\rm TS}}} {r_0^2\Gamma} \delta(r-r_0)Y_{lm}^*(\frac{\pi}{2},0) 2\pi \delta(\omega-m\Omega)\,.\qquad 
\eea
For the case under study  $\widehat {\mathcal S}_{lmn}(\omega,r)$  does not depend explicitly on $n$. 

Similarly, let us look for solutions $\psi$ in a Fourier series/integral expansion of the form
\beq
\label{psi_exp0}
\psi(t,r,\theta,\phi,y)=\sum_{lmn} \int_{\omega}e^{-i\omega t} e^{i n \frac{y}{R_y}} R_{lm n\omega}(r)Y_{lm}(\theta,\phi)\,.
\eeq

\begin{widetext}
Eq. \eqref{eq_fund} then implies
\bea
\Box \psi &=&\sum_{lmn} \int_{\omega} e^{-i \omega t}e^{in \frac{y}{R_y}}Y_{lm}(\theta,\phi)\times 
\nonumber\\
&&\left[f_s(r)f_b(r)R_{lmn\omega}''+\frac{f_s(r)+f_b(r)}{r}R_{lm\omega\omega'}' +\left(\frac{\omega^2}{f_s(r)}-\frac{n^2}{R_y^2 f_b(r)}-\frac{L}{r^2}\right)R_{lmn\omega}
\right]\nonumber\\
&=& -4\pi  \sum_{lmn}\int_{\omega}  e^{-i\omega t}e^{in \frac{y}{R_y}}\widehat {\mathcal S}_{lmn}(\omega,r) Y_{lm}(\theta, \phi)\,,
\eea
(here $L=l(l+1)$) 
that is, bringing both terms on the same side 
\beq
\sum_{lmn} Y_{lm}(\theta,\phi)\int_{\omega}e^{-i\omega t} e^{i n \frac{y}{R_y}}{\mathcal E}_{lmn\omega}(r)=0\,,
\eeq
where
\bea
{\mathcal E}_{lmn\omega}(r)&=&  f_s(r)f_b(r)R_{lmn\omega}''(r)+\frac{f_s(r)+f_b(r)}{r}R_{lm\omega\omega'}'(r)+\left(\frac{\omega^2}{f_s(r)}-\frac{n^2}{R_y^2 f_b(r)}-\frac{L}{r^2}\right)R_{lmn\omega}(r)
 +4\pi  \widehat {\mathcal S}_{lmn}(\omega,r) \,,\qquad
\eea
which will be solved  mode-by-mode by requiring
\beq
\label{eq_cal_E}
{\mathcal E}_{lmn\omega}(r)=0\,.
\eeq
Explicitly Eq. \eqref{eq_cal_E} becomes
\bea
\label{eq_rad_R_finale}
&&R_{lmn\omega}''(r)+\frac{f_s(r)+f_b(r)}{f_s(r)f_b(r) r}R_{lmn\omega}'(r)+\frac{1}{f_s(r)f_b(r)}\left(\frac{\omega^2}{f_s(r)}-\frac{n^2}{R_y^2f_b(r)}-\frac{L}{r^2}\right) R_{lmn\omega}(r)\nonumber\\
&=& -4\pi  \frac{q_{_{\rm TS}}} {r_0^2\Gamma f_s(r_0)f_b(r_0)}  \delta(r-r_0)Y_{lm}^*(\frac{\pi}{2},0) 2\pi \delta(\omega-m\Omega)
\eea
(note the relation $r(f_s(r)+f_b(r))=\Delta'$,  
where $\Delta (r)=r^2 f_s(r)f_b(r)$, Eq. \eqref{Delta_def})\,.
Eq. \eqref{eq_rad_R_finale} is conveniently rewritten by introducing the operator
\beq
\label{L_r_def0}
{\mathcal L}_{(r)}(\cdot )\equiv\frac{d^2}{dr^2}(\cdot )+\frac{f_s(r)+f_b(r)}{f_s(r)f_b(r) r}\frac{d}{dr}(\cdot )+\frac{1}{f_s(r)f_b(r)}\left(\frac{\omega^2}{f_s(r)}-\frac{n^2}{R_y^2 f_b(r)}-\frac{L}{r^2}\right)(\cdot ) \,,
\eeq
\end{widetext}
as well as the combination
\bea
S_{lmn\omega}&=&- \frac{4\pi q_{_{\rm TS}}} {\Gamma \Delta(r_0)} Y_{lm}^*(\frac{\pi}{2},0) 2\pi \delta(\omega-m\Omega)\nonumber\\
&=&\bar S_{lmn\omega}2\pi \delta(\omega-m\Omega)\,,
\eea
with
\bea
\bar S_{lmn\omega} 
&=&-q_{_{\rm TS}}K(r_0)Y_{lm}^*(\frac{\pi}{2},0)\,,
\eea
where we have also defined
\beq
K(r_0)= \frac{4\pi }{\Gamma \Delta(r_0)}\,.
\eeq
Eq. \eqref{eq_rad_R_finale} then becomes
\bea
{\mathcal L}_{(r)}R_{lmn\omega}&=& S_{lmn\omega} \delta(r-r_0)\,.
\eea
This equation can be solved by using the Green's function method, {\it i.e.} introducing the Green's function  
\beq
{\mathcal L}_{(r)} G_{lmn\omega}(r,r')=\frac{1}{\Delta(r')}\delta(r-r')\,, 
\eeq
with
\bea
G_{lmn\omega}(r,r')&=&\frac{1}{W_{lmn\omega}}[R_{\rm in}(r)R_{\rm up}(r')\Theta(r'-r)\nonumber\\
&+&R_{\rm in}(r')R_{\rm up}(r)\Theta(r-r')]\nonumber\\
&\equiv & \frac{1}{W_{lmn\omega}}R_{\rm in}(r_<)R_{\rm up}(r_>)\,,
\eea
where $r_<$ and $r_>$ correspond to $r$, $r'$ and $R_{\rm in}(r)$ and $R_{\rm up}(r)$ are two independent solutions of the homogeneous equation, both depending on $l,m,n,\omega$. They can be of various type: genuine PN, MST, JWKB, qSW/AGT solutions \cite{Bianchi:2024vmi}. Here this dependence is not shown explicitly to ease notation.

The physical range of the radial variable is $[r_b,\infty)$ and
\beq
W_{lmn\omega}=\Delta (r) [R_{\rm in}(r)R_{\rm up}'(r)-R_{\rm up}(r)R_{\rm in}'(r)]\,,
\eeq
is the constant Wronskian.
Finally,
\bea
\label{final_R}
R_{lmn\omega}(r)&=&S_{lmn\omega} \int dr' G_{lmn\omega}(r,r')\Delta(r')  \delta(r'-r_0)\nonumber\\
&=&S_{lmn\omega}G_{lmn\omega}(r,r_0) \,.
\eea
With this expression for the radial functions  we can evaluate the field
\begin{widetext}
\bea
\psi(t,r,\theta,\phi,y)&=&\sum_{lmn}\int_{\omega }e^{-i\omega t}e^{in\frac{y}{R_y}}R_{lmn\omega}(r)Y_{lm}(\theta,\phi)\nonumber\\
&=& \sum_{lmn}\int_{\omega }e^{-i\omega t}e^{in\frac{y}{R_y}}\bar S_{lmn\omega}G_{lmn\omega}(r,r_0)2\pi \delta(\omega-m\Omega) Y_{lm}(\theta,\phi)\nonumber\\
&=& \sum_{lmn} e^{in\frac{y}{R_y}}\bar S_{lmn\omega}G_{lmn\omega}(r,r_0)e^{-im\Omega t} Y_{lm}(\theta,\phi)\nonumber\\
&=& -q_{_{\rm TS}} K(r_0) \sum_{lmn} e^{in\frac{y}{R_y}} G_{lmn\omega}(r,r_0) Y_{lm}(\theta,\phi-\Omega t) Y_{lm}^*(\frac{\pi}{2},0)\,, 
\eea
which along the orbit becomes
\beq
\psi(t,r_0,\frac{\pi}{2},\Omega t,0)
= -q_{_{\rm TS}} K(r_0) \sum_{lmn}|Y_{lm}(\frac{\pi}{2},0) |^2   G_{lmn\omega}(r_0,r_0)   \,,
\eeq
{\it i.e.}   it does not depend on $t$.
\end{widetext}

In a previous study \cite{Bianchi:2024vmi}, which we will continue and complete here, we have considered the simplified situation of a string source completely delocalized in the $y$ direction, namely a \lq\lq smeared source."
This is practically obtained by taking the $y$-average of the source, Eq. \eqref{rho_def}, or equivalently to consider the $n=0$ mode only (besides the $l,m,\omega$ ones). Therefore, hereafter, of the previously generated general expressions we will only consider their $n=0$ contribution and
leave to future investigations the case with all $n\neq 0$ KK modes switched on.
The latter however correspond to fast-decaying massive waves.

The source density is then
\beq
\label{rho_def2}
\rho
=\frac{q_{_{\rm TS}}} {r_0^2\Gamma} \sum_{lm}\delta(r-r_0) Y_{lm}(\theta,\phi)e^{-im\Omega t}Y_{lm}^*(\frac{\pi}{2},0)\,,\nonumber\\
\eeq
Consequently  the field $\psi$, Eq. \eqref{psi_exp0}, turns out to satisfy a \lq\lq smeared field" condition too, corresponding to the single mode $n=0$
\beq
\label{psi_exp_semared}
\psi(t,r,\theta,\phi)=\sum_{lm} \int_{\omega}e^{-i\omega t} R_{lm\omega}(r)Y_{lm}(\theta,\phi)\,,
\eeq
and the radial equation is given by Eq. \eqref{eq_rad_R_finale} with $n=0$,
so that the field does not depend explicitly on $y$
\bea
\psi(t,r,\theta,\phi)
&=&-q_{_{\rm TS}} K(r_0) \sum_{lm} G_{lm\omega}(r,r_0)\times \nonumber\\ 
&& \times Y_{lm}(\theta,\phi-\Omega t) Y_{lm}^*(\frac{\pi}{2},0)\,,
\eea
and along the orbit it does not depend on $t$ either, 
\beq
\psi(t,r_0,\frac{\pi}{2},\Omega t)
=-q_{_{\rm TS}} K\sum_{lm}|Y_{lm}(\frac{\pi}{2},0) |^2   G_{lm\omega}(r_0,r_0)
  \,.
\eeq

In Ref. \cite{Bianchi:2024vmi} we had already reported the details of this computation. To summarize: $G_{lm\omega}(r_0,r_0)|_{\omega=m\Omega}$ is an even function of $m$, which at the accuracy we work here involves even powers of $m$ up to $m^8$ included,
\beq
G_{lm\omega}(r_0,r_0)=g_l^{(m^0)}(r_0)+g_l^{(m^2)}(r_0)m^2+g_l^{(m^4)}(r_0)m^4+\ldots
\eeq
The sum over $m$ of terms of the type
\beq
Y_l(k)=\sum_{m=-l}^l|Y_{lm}(\frac{\pi}{2},0) |^2 m^{2k}
\eeq 
is well known, see e.g. \cite{Nakano:2003he}. In fact
\bea
\Gamma_l(z)&=&\sum_{m=-l}^l e^{mz}|Y_{lm}(\frac{\pi}{2},0) |^2\nonumber\\
&=& \frac{2l+1}{4\pi}e^{lz}{}_2F_1[\frac12,-l,1,1-e^{-2z}]\nonumber\\
&=& Y_l(0)+0+\frac{z^2}{2!} Y_l(2)+\ldots\,,
\eea
i.e., explicitly
\bea
Y_l(0)&=&\frac{(1+2l)}{4\pi}\,,\nonumber\\
Y_l(2)&=&\frac{(1+2l)}{4\pi} \frac{l (l+1)}{2}\,,\nonumber\\
Y_l(4)&=&\frac{(1+2l)}{4\pi} \frac{l (l+1) (3 l^2+3 l-2)}{8}\,,\nonumber\\
Y_l(6)&=&\frac{(1+2l)}{4\pi} \frac{l (l+1) (5 l^4+10 l^3-5 l^2-10 l+8)}{16}\,,\nonumber\\
Y_l(8)&=& \frac{(1+2l)}{4\pi} \frac{ l (l+1)}{128} (35 l^6+105 l^5-35 l^4\nonumber\\
&&-245 l^3+168 l^2+308 l-272)  \,.
\eea

Therefore
\bea
\sum_{lm}|Y_{lm}(\frac{\pi}{2},0) |^2   G_{lm\omega}(r_0,r_0)&=& 
g_l^{(m^0)}(r_0)Y_l(0)\nonumber\\
&+& g_l^{(m^2)}(r_0)Y_l(2)\nonumber\\
&+& g_l^{(m^4)}(r_0)Y_l(4)\nonumber\\
&+& \ldots
\eea
At this point, in principle, the sum over $l$ can be performed. However, it diverges, and according to a well established procedure, one has to subtract the singular field (see e.g., \cite{Detweiler:2002gi})
\beq
B=\lim_{l\to \infty} \sum_{k=0}^\infty g_l^{(m^k)}(r_0)Y_l(k)\,.
\eeq
After this subtraction (regularization) process one is left with an expression which cannot be summed over $l$ from zero to infinity at the accuracy we are working with.
 In the post-Newtonian (PN) expansion in $\eta_v=v/c$, in order to achieve $O(\eta_v^9)$ accuracy, we sum from $l=3$ to infinity by using the PN solution, while  for the $l=0,1,2$ terms we use the MST solutions. 
This leads to the result accounted in Ref. \cite{Bianchi:2024vmi}. Useful details are included in the Supplemental Material associated with the present work.

\subsection{Self-force on the particle's orbit}

Let us recall the definition of the self-force in 4D
\beq
\label{4D_force}
F^\mu =q {\cal P}({{U}})^{\mu\nu}\partial_\nu \psi\,,
\eeq
where ${\cal P}({{U}})=g+{{U}}\otimes {{U}}$ projects orthogonally to ${{U}}$, the particle's four velocity. 
For dimensional reasons, the 5D transcription of Eq. \eqref{4D_force} is  simply 
\beq
F^\mu =q_{_{\rm TS}} {\cal P}({{U}})^{\mu\nu}\partial_\nu \psi\,.
\eeq
Because of the dependence on the combination $\phi-\Omega t$ of the field (and not on $t$ and $\phi$ separately), we have ${{U}}^\mu \partial_\mu \psi=0$, and then simply
\bea
F_\mu &=& q_{_{\rm TS}} \partial_\mu \psi\nonumber\\
&=& -q_{_{\rm TS}}^2  K(r_0) \sum_{lm}Y_{lm}^*(\frac{\pi}{2}, 0)  \partial_\mu \,\Big[ Y_{lm}(\theta,\phi-\Omega t) \nonumber\\
&& \times  G_{lm\omega}(r,r_0)\bigg|_{\omega=m\Omega}\Big]\,,
\eea
with 
\beq 
F_\phi=-\frac{F_t}{\Omega}\,,
\eeq
since $\psi$ depends only on the combination $\phi-\Omega t$, and 
\beq
F_y=0\,,
\eeq
since $\psi$ is independent of $y$, thanks to `smearing' or continuous shift-invariance (for a small string) under $y\rightarrow y+ \beta$.
A direct evaluation of the self force along the orbit gives
\beq
F_\theta|^{\rm orb} =0\,.
\eeq
In fact 
\bea
&&\left[Y_{lm}^*(\frac{\pi}{2}, 0)  \partial_\theta \,  Y_{lm}(\theta,\phi-\Omega t)\right]_{\phi=\Omega t, \theta=\pi/2}  =  \nonumber \\  
&&  \frac{2l+1}{4\pi \Gamma(-l-m)\Gamma(l+m+1)}=0\,,
\eea
since $l+m=$ integer.
We are then left with 
\bea
F_t|^{\rm orb}  
&=& -q_{_{\rm TS}}^2  K \sum_{lm}|Y_{lm}^*(\frac{\pi}{2}, 0)|^2  (-im\Omega) \,  G_{lm\omega}(r_0,r_0)\,,
\nonumber\\
F_r|^{\rm orb}  
&=& -q_{_{\rm TS}}^2  K \sum_{lm}|Y_{lm}^*(\frac{\pi}{2}, 0)|^2  \partial_r G_{lm\omega}(r,r_0)|_{r=r_0}
\,,\nonumber\\
\eea
with $\omega=m\Omega$ implicitly understood whenever going along the orbit. 
Note that when computing explicitly the above sums one should regularize them, since they are generally divergent. Moreover,
\beq
G_{lm\omega}(r,r_0)=\left\{
\begin{array}{cc}
\frac{1}{W_{lm\omega}}R_{\rm in}(r)R_{\rm up}(r_0)&\quad r<r_0\,, \hbox{left}, -\cr
& \cr
\frac{1}{W_{lm\omega}}R_{\rm in}(r_0)R_{\rm up}(r)&\quad r>r_0\,, \hbox{right}, +\cr
\end{array}
\right.
\eeq
and consequently we have both a left force and a right force  for each component,
\bea
F_t|^{{\rm orb}, -}  
&=& -q_{_{\rm TS}}^2  K \sum_{lm}|Y_{lm}^*(\frac{\pi}{2}, 0)|^2  (-im\Omega) \, G_{lm\omega}(r_0,r_0)\nonumber\\
&=& F_t|^{{\rm orb}, +}
\nonumber\\
F_r|^{{\rm orb},-}  
&=& -q_{_{\rm TS}}^2  K\sum_{lm}|Y_{lm}^*(\frac{\pi}{2}, 0)|^2 H_{lm\omega}^-(r_0)
\nonumber\\
F_r|^{{\rm orb},+}  
&=& -q_{_{\rm TS}}^2  K\sum_{lm}|Y_{lm}^*(\frac{\pi}{2}, 0)|^2 H_{lm\omega}^+(r_0)
\,,
\eea
where we have defined
\bea
H_{lm\omega}^-(r_0)&=&\frac{R_{\rm in}'(r_0)R_{\rm up}(r_0)}{W_{lm\omega}}\,,\nonumber\\
H_{lm\omega}^+(r_0)&=& \frac{R_{\rm up}'(r_0)R_{\rm in}(r_0)}{W_{lm\omega}}\,.
\eea
A direct computation actually shows 
\beq
F_t|^{{\rm orb}, \pm}=0\,,
\eeq 
because $G_{lm\omega}(r_0,r_0)$ is an even polynomial in $m$ which multiplied by $m\Omega$ becomes odd and sums to zero \cite{Nakano:2003he}.
When computing these quantities it is worth to recall  the  PN  relation
\beq
R_{\rm up}(r_0)=R_{\rm in}(r_0)\big|_{l\to -l-1}\,,
\eeq
which implies, for example, a simple difference $\Delta H=H_{lm\omega}^+(r_0)-H_{lm\omega}^-(r_0)$ not depending on either $l,m$
\bea
\Delta H&=& \frac{4u^2}{r_s^2}\left[1+2u(1+\alpha)+(2u)^2(1+\alpha+\alpha^2)\right.\nonumber\\
&+&  (2u)^3(1+\alpha+\alpha^2+\alpha^3)\nonumber\\
&+&\left. (2u)^4(1+\alpha+\alpha^2+\alpha^3+\alpha^4)  \right]\nonumber\\
&=& \frac{4u^2}{r_s^2} \sum_{k=0}^\infty (2u)^k \sum_{i=0}^k \alpha^i\nonumber\\
&=& \frac{4u^2}{r_s^2} \frac{1}{(1-2u)(1-2\alpha u)}\,.
\eea
This implies a jump in $F_r|^{{\rm orb}, \pm}$ which the requires average to be regularized.
We list below the (averaged, regularized) radial component  of the force up to the approximation level considered here, 
\bea
\label{force_terms}
F_r^{l=0} &=&-\frac{u^3}{8 \pi  r_s^2}\left(\tilde F^{\alpha^0}_{l=0}+\alpha \tilde F^{\alpha^1}_{l=0}+\alpha^2 \tilde F^{\alpha^2}_{l=0}\right.\nonumber\\
&+&\left. \alpha^3 \tilde F^{\alpha^3}_{l=0}+\alpha^4 \tilde F^{\alpha^4}_{l=0}\right)\,,\nonumber\\
F_r^{l=1} &=&-\frac{3u^3}{40 \pi r_s^2}\left(\tilde F^{\alpha^0}_{l=1}+\alpha \tilde F^{\alpha^1}_{l=1}+ \alpha^2 \tilde F^{\alpha^2}_{l=1}\right.\nonumber\\
&+&\left. \alpha^3 \tilde F^{\alpha^3}_{l=1}+\alpha^4 \tilde F^{\alpha^4}_{l=1}\right)\,,\nonumber\\
F_r^{l=2} &=&-\frac{u^3}{56 \pi r_s^2}\left(\tilde F^{\alpha^0}_{l=2}+\alpha \tilde F^{\alpha^1}_{l=2}+ \alpha^2 \tilde F^{\alpha^2}_{l=2}\right.\nonumber\\
&+&\left. \alpha^3 \tilde F^{\alpha^3}_{l=2}+\alpha^4 \tilde F^{\alpha^4}_{l=2}\right)\,,\nonumber\\
\sum_{l=3}^\infty F_r  &=&-\frac{9 u^3}{280 \pi r_s^2}\left(\tilde F^{\alpha^0}_{\Sigma_{l\ge 3}}+\alpha \tilde F^{\alpha^1}_{\Sigma_{l\ge 3}}+ \alpha^2 \tilde F^{\alpha^2}_{\Sigma_{l\ge 3}}\right.\nonumber\\
&+& \left. \alpha^3 \tilde F^{\alpha^3}_{\Sigma_{l\ge 3}}+\alpha^4 \tilde F^{\alpha^4}_{\Sigma_{l\ge 3}}\right)\,,
\eea
with all the various terms listed in Table \ref{tab:1} below.
\begin{widetext}
\begin{table*}  
\caption{\label{tab:1}  Contributions to the various force terms entering Eq. \eqref{force_terms}.}
\begin{ruledtabular}
\begin{tabular}{ll}
$\tilde F^{\alpha^0}_{l=0} $ & $ \frac{145921}{4096}u^3+\frac{507}{64}u^2+\frac{5}{16}u-1$\\
$\tilde F^{\alpha^1}_{l=0}$ & $\frac{27}{8}u^3+\frac{49}{8}u^2+4u$\\ 
$\tilde F^{\alpha^2}_{l=0}$ & $\frac{347 }{32}u^3+\frac{13}{2}u^2-2u$\\
$\tilde F^{\alpha^3}_{l=0}$ & $9u^3-8u^2 $\\
$\tilde F^{\alpha^4}_{l=0}$ &$ -\frac{47}{2}u^2$\\
\hline
$\tilde F^{\alpha^0}_{l=1}$ & $-\frac{304}{27} \pi  u^{7/2}+u^3 \left(\frac{112 \log
   (u)}{9}+\frac{224 \gamma
   }{9}-\frac{19917478621}{638668800}+\frac{224 \log
   (2)}{9}\right)+u^2 \left(-\frac{80 \log (u)}{9}-\frac{160
   \gamma }{9}-\frac{253513}{181440}-\frac{160 \log
   (2)}{9}\right)-\frac{923 u}{112}+1$\\
$\tilde F^{\alpha^1}_{l=1}$ & $-\frac{112}{27} \pi  u^{7/2}+u^3 \left(\frac{136 \log
   (u)}{9}+\frac{272 \gamma
   }{9}+\frac{3600781}{113400}+\frac{272 \log
   (2)}{9}\right)+u^2 \left(-\frac{40 \log (u)}{9}-\frac{80
   \gamma }{9}+\frac{5707}{280}-\frac{80 \log
   (2)}{9}\right)+4 u$\\
$\tilde F^{\alpha^2}_{l=1}$ & $-\frac{64}{27} \pi  u^{7/2}+u^3 \left(\frac{40 \log
   (u)}{9}+\frac{80 \gamma
   }{9}+\frac{5969071}{151200}+\frac{80 \log
   (2)}{9}\right)+\frac{2051 u^2}{270}-2 u$\\
$\tilde F^{\alpha^3}_{l=1}$ & $\frac{4097 u^3}{945}-8 u^2$\\
$\tilde F^{\alpha^4}_{l=1}$ & $-\frac{305 u^3}{14}$\\
\hline
$ \tilde F^{\alpha^0}_{l=2}$ & $u^3 \left(-\frac{3584 \log (u)}{15}-\frac{7168 \gamma
   }{15}+\frac{14163237695083}{19372953600}-\frac{14336 \log
   (2)}{15}\right)-\frac{108072103 u^2}{1108800}+\frac{291
   u}{16}+1$\\
$\tilde F^{\alpha^1}_{l=2}$ & $u^3 \left(-\frac{1792 \log (u)}{15}-\frac{3584 \gamma
   }{15}+\frac{561682481}{970200}-\frac{7168 \log
   (2)}{15}\right)-\frac{69803 u^2}{4200}+4 u$\\
$\tilde F^{\alpha^2}_{l=2}$ & $\frac{1404083 u^3}{352800}+\frac{8273 u^2}{1050}-2 u$\\
$\tilde F^{\alpha^3}_{l=2}$ & $\frac{14573 u^3}{525}-8 u^2$\\
$\tilde F^{\alpha^4}_{l=2}$ & $-\frac{55 u^3}{2}$\\
\hline
$\tilde F^{\alpha^0}_{\Sigma_{l\ge 3}}$ & $\left(\frac{6895 \pi
   ^2}{1152}-\frac{11284849260613}{174356582400}\right)
   u^3+\left(\frac{23576467}{1197504}+\frac{245 \pi
   ^2}{72}\right) u^2+\frac{1139 u}{144}+1$\\
$\tilde F^{\alpha^1}_{\Sigma_{l\ge 3}}$ & $\left(\frac{5845 \pi
   ^2}{576}-\frac{4643454749}{26195400}\right)
   u^3+\left(\frac{35 \pi
   ^2}{18}-\frac{407389}{22680}\right) u^2+4 u$\\
$\tilde F^{\alpha^2}_{l\ge 3}$ & $\left(\frac{5495 \pi
   ^2}{1152}-\frac{37823423}{1058400}\right)
   u^3+\left(\frac{18007}{5670}+\frac{35 \pi ^2}{72}\right)
   u^2-2 u$\\
$\tilde F^{\alpha^3}_{\Sigma_{l\ge 3}}$ & $\left(\frac{26827}{2835}+\frac{35 \pi ^2}{36}\right) u^3-8
   u^2$\\
$\tilde F^{\alpha^4}_{\Sigma_{l\ge 3}}$ & $-\frac{455 u^3}{18}$\\
 \end{tabular}
\end{ruledtabular}
\end{table*}

Summing all contributions we find
\bea
F_r^{\rm tot} &=& \tilde F^{\alpha^0}_{\rm tot}+\alpha \tilde F^{\alpha^1}_{\rm tot}+ \alpha^2 \tilde F^{\alpha^2}_{\rm tot}+\alpha^3 \tilde F^{\alpha^3}_{\rm tot} \,,
\eea
with
\bea
\tilde F^{\alpha^0}_{\rm tot}&=&\frac{1}{576 \pi r_s^2}\left[\frac{2432}{5} \pi  u^{13/2}+u^6 \left(1920 \log
    u -\frac{1773 \pi ^2}{16}+3840 \gamma+\frac{43776}{5} \log 2-\frac{37696}{5}\right)\right.\nonumber\\
&+&\left. u^5 \left(384 \log (u)-63 \pi
   ^2+768 \gamma +128+768 \log  2 \right) 
\right] \nonumber\\ 
\tilde F^{\alpha^1}_{\rm tot}&=& \frac{14 u^{13/2}}{45 r_s^2}+\frac{u^6}{\pi  r_s^2}
\left( \log
    u  -\frac{33}{5}+2\gamma -\frac{167\pi^2}{512}+\frac{94}{15}\log 2\right)
\nonumber\\
&+&\frac{u^5}{\pi 
   r_s^2} \left( \frac13 \log u +\frac19 -\frac{\pi ^2}{16}+\frac23  \gamma+\frac23  \log  2 \right) \nonumber\\ 
\tilde F^{\alpha^2}_{\rm tot}&=& \frac{8 u^{13/2}}{45 r_s^2}-\frac{u^6}{\pi r_s^2} \left(\frac13\log
    u +\frac{19}{36}+\frac{157\pi^2}{1024}+\frac23 \gamma +\frac23 \log 2\right) -\frac{\pi  u^5}{64 r_s^2}\nonumber\\ 
\tilde F^{\alpha^3}_{\rm tot}&=& -\frac{\pi  u^6}{32 r_s^2}\,.
\eea

\end{widetext}
 
A consequence of the present analysis is that the self-force along the world line of the source particle is a constant force, directed only radially. We are then motivated to study the modifications to the circular orbit under the effect of a (generic) constant force on a test particle in the TS background.  Notice that as soon as the corrections to circular motion due to the presence of the external force is taken into account, at higher orders in the $q$-expansion the force itself would not be  a constant anymore, and motion would not be\lq\lq circular." However, we work here only at the first order in $q$ and hence the force can be considered  as a constant.

\section{Equations of motion and orbit modified by a constant force in a TS background}

The orbit of a charged particle, with four velocity $U$, is accelerated (with acceleration $a(U)=\nabla_U U$) by  a force $q F$, and satisfies the standard relation $m a(U) =q F$, i.e., 
\beq
\frac{d{{U}}^\lambda}{d\tau}+\Gamma^{\lambda}{}_{\mu\nu}{{U}}^\mu {{U}}^\nu =\frac{q}{m}F^\lambda\,,
\eeq
with ${{U}}^\mu \partial_\mu\psi=0$,
and the force $F$ which causes deviations from circularity~\footnote{For the moment $qF$ is a generic constant force. Later we will consider the case in which it coincides with the self force.}.
Let us write
\beq
{{U}}^\mu(\tau)={{U}}^\mu_{\rm geo, circ}(\tau)+q\delta {{U}}^\mu(\tau)\,,
\eeq
with corresponding parametric equations $x^\mu(\tau)=x^\mu_{\rm geo, circ}(\tau)+q\delta x^\mu(\tau)$, namely
\bea
t&=&\frac{\tau}{\sqrt{f_{\rm ph}(r_0)}}+q\delta t(\tau)\,,\nonumber\\ 
r&=& r_0+q\delta r(\tau)\,,\nonumber\\ 
\theta&=&\frac{\pi}{2}+q\delta \theta(\tau)\,,\nonumber\\ 
\phi &=&\frac{\Omega\tau}{\sqrt{f_{\rm ph}(r_0)}}+q\delta \phi(\tau)\,,\nonumber\\
y&=&q\delta y(\tau)\,,
\eea
so that $\delta{{U}}^\mu=\frac{d \delta x^\mu}{d\tau}$.
Note that up to now  we have assumed the force with all components, namely in principle to be more general than the self-force expressions derived in the previous section.
Soon we will make additional assumptions.

The normalization condition for $U$ timelike, $U\cdot U=-1$, implies
\beq
\delta{{U}}^t(\tau) =  \frac{r_0^2\Omega }{ f_s(r_0)}\delta{{U}}^\phi(\tau)\,,
\eeq
while the solution of the equations of motion, after imposing equatorial motion also in the perturbed case or $F^\theta=0$ and $\delta{{U}}^\theta=0$ (so that we can assume $\delta \theta\equiv 0$), reads
\bea
\delta{{U}}^t(\tau) &=&  C_1^t\tau+C_2^t\sin(\Omega_0\tau)+C_3^t(1-\cos(\Omega_0\tau))\nonumber\\
&+& \delta U^t(0), \nonumber\\
\delta{{U}}^r(\tau) &=& C_2^r(1-\cos(\Omega_0\tau))+C_3^r\sin(\Omega_0\tau)\nonumber\\
&+& \delta U^r(0),\nonumber\\ 
\delta{{U}}^\phi(\tau) &=&C_1^\phi\tau +C_2^\phi\sin(\Omega_0\tau)+C_3^\phi(1-\cos(\Omega_0\tau))\nonumber\\
&+& \delta U^\phi(0),\nonumber\\
\delta{{U}}^y(\tau) &=& \frac{F^y}{m}\tau+\delta U^y(0)\,, 
\eea
with
\bea
\delta t(\tau)  &=& \frac12 C_1^t\tau^2-\frac{C_2^t}{\Omega_0}\cos(\Omega_0\tau)+C_3^t(\tau -\frac{1}{\Omega_0}\sin(\Omega_0\tau))\nonumber\\
&+& \delta{{U}}^t(0)\tau +C_4^t\,, \nonumber\\
\delta r(\tau)   &=& \delta{{U}}^r(0)\tau+C_2^r(\tau-\frac{1}{\Omega_0}\sin(\Omega_0\tau))-\frac{C_3^r}{\Omega_0}\cos(\Omega_0\tau)\nonumber\\
&+& C_4^r\,, \nonumber\\
\delta \phi(\tau)&=& \frac12 C_1^\phi\tau^2 -\frac{C_2^\phi}{\Omega_0}\cos(\Omega_0\tau)+C_3^\phi(\tau-\frac{1}{\Omega_0}\sin(\Omega_0\tau))\nonumber\\
&+& \delta{{U}}^\phi(0)\tau +C_4^\phi\,, \nonumber\\
\delta y(\tau)&=& \frac12 \frac{F^y}{m}\tau^2+\delta{{U}}^y(0)\tau+\delta y(0)\,,
\eea
with the constraint, due to the orthogonality between $F$ (order $q$) and ${U}$ (order $0$)
\beq
f_s(r_0) F^t=\Omega  r_0^2 F^\phi\,,\qquad - F_t=\Omega F_\phi\,,  
\eeq
and from the normalization condition
\beq
 C_i^t= \frac{r_0^2\Omega }{ f_s(r_0)} C_i^\phi\,,\quad i=1,\ldots, 4\,,
\eeq
and,  denoting henceforth $f_s=f_s(r_0)$, $f_b=f_b(r_0)$, 
\bea
\label{int_coonst_gen}
C_2^r &=&\frac{2 f_s f_{\rm ph}^{3/2}}{ r_0 f_{_{\rm ISCO}}\Omega^2} \frac{F^t}{m}-\delta U^r(0), \nonumber\\
C_3^r &=& \frac{3 f_b^{1/2} f_{_{\rm ISCO}}^{1/2} f_s\Omega }{ 4f_{\rm ph}^{3/2}}C_4^r
+\frac{f_{_{\rm ISCO}}^{1/2} }{ 4 f_b^{1/2} f_{\rm ph}^{1/2}\Omega}\frac{F^r}{m}\nonumber\\
&+& \frac{r_0 f_b^{1/2}f_{_{\rm ISCO}}^{1/2}}{ 2} \delta U^\phi(0),\nonumber\\ 
C_1^\phi &=& \frac{(1-4 \frac{f_{\rm ph}}{f_{_{\rm ISCO}}}) f_s}{ r_0^2\Omega} \frac{F^t}{m}, \nonumber\\
C_2^\phi &=& \frac{4 f_s f_{\rm ph}^{3/2}}{ r_0^2 f_{_{\rm ISCO}}^{3/2} f_b^{1/2}\Omega^2} \frac{F^t}{m}
-\frac{2}{ r_0 f_b^{1/2} f_{_{\rm ISCO}}^{1/2}}\delta U^r(0),\nonumber\\ 
C_3^\phi &=& -\frac{1}{2 r_0 f_b f_{\rm ph}^{1/2}\Omega}  \frac{F^r}{m}-\frac{3 f_s\Omega}{ 2 r_0 f_{\rm ph}^{3/2}} C_4^r-\delta U^\phi(0)\,,
\eea
with
\beq
\Omega_0=\Omega \sqrt{\frac{f_b(r_0)f_{_{\rm ISCO}}(r_0)}{f_{\rm ph}(r_0)}}\,,\qquad \Omega=\sqrt{\frac{M}{r_0^3}}\,.
\eeq
$\Omega_0$ generalizes (due to the presence of $r_b$) the epicyclic frequency, often used in modeling accretion disc particle motion around BHs.
Because of the $f_{_{\rm ISCO}}(r_0)$ entering $\Omega_0$  we assume $r_0>3r_s=6M$.  For convenience, when needed we will also measure $r_b$ in units of $r_s$, i.e.  
$r_b=\alpha r_s$. 

A special choice of initial conditions is used to simplify these expressions.
We can assume
\beq
\delta{{U}}^\mu(0)=0=\delta x^\mu(0) \,,
\eeq
implying  
\beq
C_4^t=  \frac{C_2^t}{\Omega_0}\,,\qquad C_4^r =\frac{C_3^r}{\Omega_0} \,,\qquad
C_4^\phi= \frac{C_2^\phi}{\Omega_0}\,.
\eeq
We  then have
\bea
\delta{{U}}^t(\tau) &=&  C_1^t\tau+C_2^t\sin(\Omega_0\tau)+C_3^t(1-\cos(\Omega_0\tau)), \nonumber\\
\delta{{U}}^r(\tau) &=& C_2^r(1-\cos(\Omega_0\tau))+C_3^r\sin(\Omega_0\tau),\nonumber\\ 
\delta{{U}}^\phi(\tau) &=&C_1^\phi\tau +C_2^\phi\sin(\Omega_0\tau)+C_3^\phi(1-\cos(\Omega_0\tau)),\nonumber\\
\delta{{U}}^y(\tau) &=& \frac{F^y}{m}\tau, 
\eea
with
\bea
\label{fin_per_plot}
\delta t(\tau)  &=& \frac12 C_1^t\tau^2+\frac{C_2^t}{\Omega_0}(1-\cos(\Omega_0\tau))\nonumber\\
&+&C_3^t(\tau -\frac{1}{\Omega_0}\sin(\Omega_0\tau)) \,,\nonumber\\
\delta r(\tau)   &=&  C_2^r(\tau-\frac{1}{\Omega_0}\sin(\Omega_0\tau))+\frac{C_3^r}{\Omega_0}(1-\cos(\Omega_0\tau))\,, \nonumber\\
\delta \phi(\tau)&=& \frac12 C_1^\phi\tau^2 +\frac{C_2^\phi}{\Omega_0}(1-\cos(\Omega_0\tau))\nonumber\\
&+&C_3^\phi(\tau-\frac{1}{\Omega_0}\sin(\Omega_0\tau))\,,\nonumber\\
\delta y(\tau)&=& \frac12 \frac{F^y}{m}\tau^2\,,
\eea
and then Eqs. \eqref{int_coonst_gen} reduce to
\bea
C_2^r &=&\frac{2 f_s f_{\rm ph}^{3/2}}{ r_0\Omega^2 f_{_{\rm ISCO}}} \frac{F^t}{m}\,,\qquad  
C_3^r=  \frac{ f_{\rm ph}^{1/2} }{\Omega  f_{_{\rm ISCO}}^{1/2}f_b^{1/2}}\frac{F^r}{m}
\,,\nonumber\\ 
C_1^\phi &=& -\frac{3f_s^2}{ r_0^2\Omega f_{_{\rm ISCO}}}    \frac{F^t}{m}\,,\qquad  
C_2^\phi = \frac{4 f_s f_{\rm ph}^{3/2}}{ r_0^2\Omega^2 f_{_{\rm ISCO}}^{3/2} f_b^{1/2}} \frac{F^t}{m}\,,\nonumber\\
C_3^\phi &=&  -\frac{2 f_{\rm ph}^{1/2}}{ r_0\Omega f_b f_{_{\rm ISCO}}}\frac{F^r}{m}\,,
\eea
that is
\bea
[C_3^r,C_3^\phi]&=&\frac{F^r}{m} \left[\Omega_0, -\frac{2}{r_0f_{\rm ph}^{1/2}}  \right]\,, \nonumber\\
{}[C_2^r,C_1^\phi, C_2^\phi] &=& \frac{F^t}{m}\frac{f_s}{\Omega r_0}\left[\frac{2f_{\rm ph}^{3/2}}{\Omega}, -\frac{3f_s}{r_0}, \frac{4f_{\rm ph}}{\Omega_0 r_0}  \right]\,.
\eea 

\begin{figure}
\includegraphics[scale=0.3]{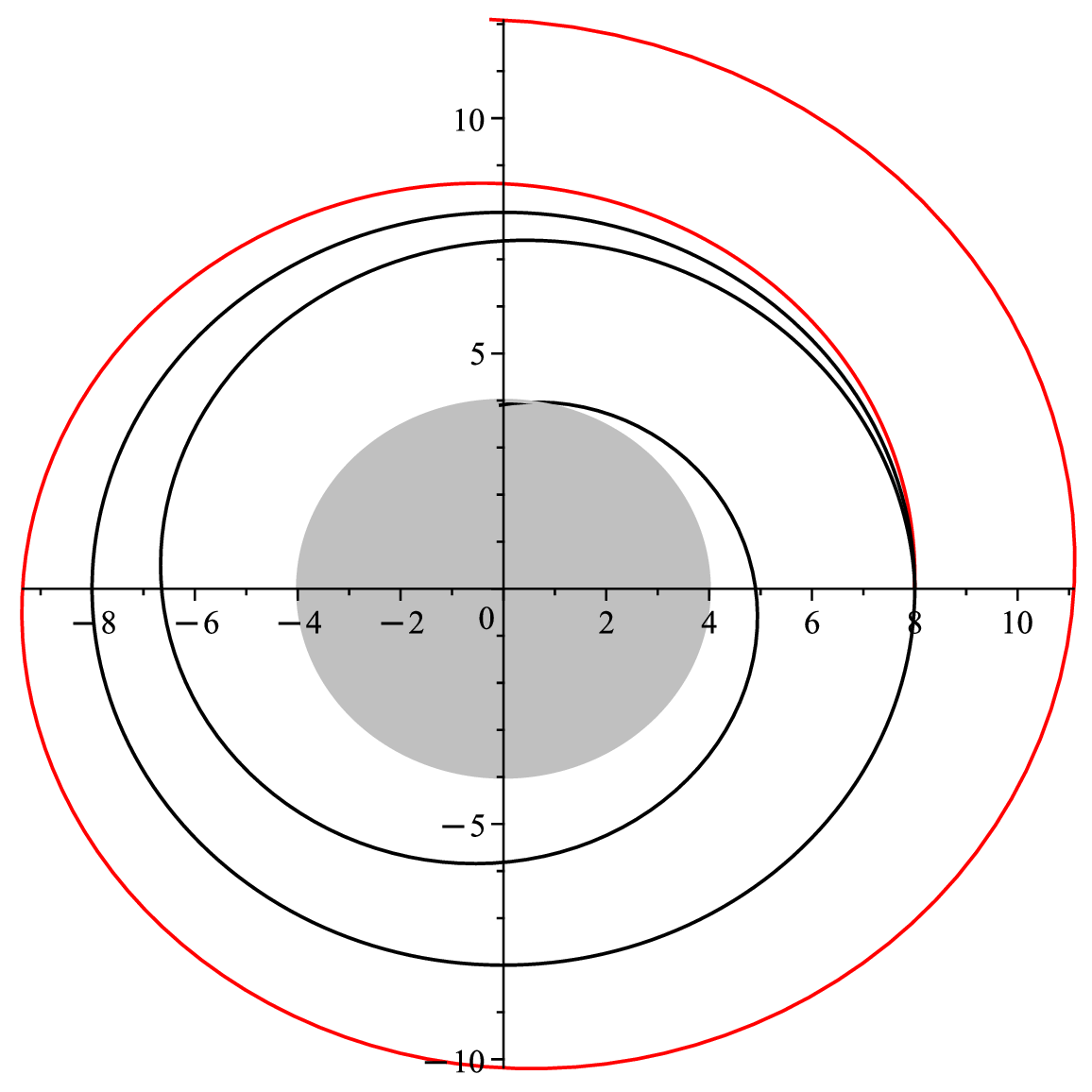}
\caption{\label{fig:1} Example of orbit  undergoing a constant force in the equatorial plane $\theta=\pi/2$, with Cartesian-like coordinates $X=r(\tau)\cos\phi(\tau)$ and $Y=r(\tau)\sin\phi(\tau)$ obtained by using Eq. \eqref{fin_per_plot}, deviating from the corresponding circular orbit (dotted curve). We have chosen
$r_0=8$, $r_b=18/5$, $M=1$, $F^t/m=F^r/m=1$ and $q=-0.1$ (black on-line) and $q=+0.1$ (red on-line). The plot of the (black  on-line) orbit stops at $\tau=8.2$ when it reaches the circle of radius $r_b$ (boundary of grey-filled region).
}
\end{figure}

Assuming $F^y=0$ (as from the self-force analysis of the previous section) implies $\delta{{U}}^y(\tau) =0=\delta y(\tau)$ identically, i.e., trivializes the role of the extra dimension $y$ and reinforces the \lq\lq smeared orbit approximation" adopted here.
Assuming further $F^t=0$ (i.e., $C_2^r=C_1^\phi=C_2^\phi=0$) we obtain (besides $\delta {{U}}^y(\tau)=0, \delta y(\tau)=0$)
\bea
\delta{{U}}^t(\tau) &=& \frac{r_0^2\Omega }{ f_s(r_0)} C_3^\phi (1-\cos(\Omega_0\tau)), \nonumber\\
\delta{{U}}^r(\tau) &=&  C_3^r\sin(\Omega_0\tau),\nonumber\\ 
\delta{{U}}^\phi(\tau) &=& C_3^\phi(1-\cos(\Omega_0\tau)),
\eea
with
\bea
\label{fin_per_plot}
\delta t(\tau)  &=&  \frac{r_0^2\Omega }{ f_s(r_0)} C_3^\phi (\tau -\frac{1}{\Omega_0}\sin(\Omega_0\tau)) \,,\nonumber\\
\delta r(\tau)   &=&   \frac{C_3^r}{\Omega_0}(1-\cos(\Omega_0\tau))\,, \nonumber\\
\delta \phi(\tau)&=&  C_3^\phi(\tau-\frac{1}{\Omega_0}\sin(\Omega_0\tau))\,.
\eea
This final, simplified case corresponds to the self-force one where we have shown that only the radial component of the force survives.

\subsection{Radiated energy and angular momentum}

In the following, we will consider the loss of energy and angular momentum by emission of (massless) scalar waves. The formulae for the radiated energy and angular momentum can be derived from the associated energy-momentum tensor which  
for a massless complex scalar field is given by
\beq
8\pi T^{\rm scal}_{\mu\nu}=\partial_\mu\psi^*\partial_\nu\psi+ \partial_\mu\psi \partial_\nu\psi^* -  g_{\mu\nu} \partial_\lambda \psi^* \partial^\lambda\psi \,,
\eeq
so that
\beq
\frac{d^2E}{dtd\Omega}=\lim_{r\to \infty} (r^2 T^{\rm scal}{}^r{}_t)\,.
\eeq
We find
\beq
8\pi T^{\rm scal}_{rt}=\psi^*_{,r}\psi_{,t}+\psi_{,r}\psi^*_{,t} \,,
\eeq
implying
\bea
8\pi T^{\rm scal}{}^r{}_{t}&=& f_sf_b (\psi^*_{,r}\psi_{,t}+\psi_{,r}\psi^*_{,t})\nonumber\\
&=& f_sf_b \psi^*_{,r}\psi_{,t}+{\rm c.c.}\,.
\eea
Here, suppressing the dependence on the spacetime variables to ease notation and assuming only emission of massless $n=0$ KK modes (higher KK would be massive from a $d=4$ perspective),
\bea
\psi&=& \sum_{lm}Y_{lm}\int \frac{d\omega}{2\pi}e^{-i\omega t}R_{lm\omega}\,,\nonumber\\
\psi_{,t}&=& \sum_{lm}Y_{lm}\int \frac{d\omega}{2\pi}(-i\omega) e^{-i\omega t}R_{lm\omega}\,,\nonumber\\
\psi^*_{,r}&=& \sum_{l'm'}Y_{l'm'}^*\int \frac{d\omega'}{2\pi}e^{i\omega' t}\frac{d}{dr}R_{l'm'\omega'}^*\,,
\eea 
and therefore
\begin{widetext}
\bea
r^2 T^{\rm scal}{}^r{}_{t}&=&\frac{r^2 f_sf_b}{8\pi} (\psi^*_{,r}\psi_{,t}+\psi_{,r}\psi^*_{,t})\nonumber\\
&=& \frac{r^2 f_sf_b}{8\pi} \sum_{lm,l'm'}\int \frac{d\omega}{2\pi} \frac{d\omega'}{2\pi}Y_{lm}(\theta,\phi)Y_{l'm'}^*(\theta,\phi)  \left[(-i\omega) e^{-i(\omega-\omega')t} R_{lm\omega}(r)\frac{d}{dr}R^*_{l'm'\omega'}(r)\right.\nonumber\\
&+& \left. (i\omega')e^{i(\omega-\omega')t} R^*_{l'm'\omega'}(r)\frac{d}{dr}R_{l m \omega }(r)\right]\,.
\eea
\end{widetext}

\hbox{\vspace{2cm}}

Integrating over the sphere and recalling the orthogonality property 
\beq
\int \sin \theta d\theta d\phi   Y_{lm}^*(\theta,\phi)Y_{l'm'}(\theta,\phi)=\delta_{ll'}\delta_{m m'}\,,
\eeq
we find
\bea
\frac{dE}{dt}&=&\lim_{r\to \infty}\int \sin \theta d\theta d\phi r^2 T^{\rm scal}{}^r{}_{t}\nonumber\\
&=&\lim_{r\to \infty} \frac{\Delta (r)}{8\pi} \sum_{lm}\int \frac{d\omega}{2\pi} \frac{d\omega'}{2\pi}\Big[ \nonumber\\
&&(-i\omega) e^{-i(\omega-\omega')t} R_{lm\omega}(r)\frac{d}{dr}R^*_{lm\omega'}(r)\nonumber\\
&+&  (i\omega')e^{i(\omega-\omega')t} R^*_{lm\omega'}(r)\frac{d}{dr}R_{l m \omega }(r)\Big]\,.
\eea

Because of the $r\to \infty$ limit we can use here only the up-part of the $R_{lm\omega}(r)$ solution, namely
\bea
\label{sol_con_Z}
R_{lm\omega}(r)
&=& \frac{R_{\rm up}(r)}{W_{lm\omega}}  R_{\rm in}(r_0) \Delta (r_0)\bar S_{lm\omega}2\pi \delta(\omega-m\Omega)\nonumber\\
&=& {\mathfrak R}_{lm\omega}(r)2\pi \delta(\omega-m\Omega)
\,,
\eea
with
\bea
 {\mathfrak R}_{lm\omega}(r)&=&- \frac{4\pi q_{_{\rm TS}}} {\Gamma}\frac{R_{\rm in}(r_0)  Y_{lm}^*(\frac{\pi}{2},0)}{W_{lm\omega}}  R_{\rm up}(r)\nonumber\\
&=&q_{_{\rm TS}} C_{lm\omega}(r_0)R_{\rm up}(r)\,,
\eea
with
\beq
 C_{lm\omega}(r_0)=- \frac{4\pi} {\Gamma}\frac{R_{\rm in}(r_0)  Y_{lm}^*(\frac{\pi}{2},0)}{W_{lm\omega}}\,.
\eeq

Finally,
\begin{widetext}
\bea
\frac{dE}{dt}
&=&\lim_{r\to \infty} \frac{\Delta(r)}{8\pi} \sum_{lm}\int \frac{d\omega}{2\pi} \frac{d\omega'}{2\pi}\left[(-i\omega) e^{-i(\omega-\omega')t} R_{lm\omega}(r)\frac{d}{dr}R^*_{lm\omega'}(r)\right.\nonumber\\
&+& \left. (i\omega')e^{i(\omega-\omega')t} R^*_{lm\omega'}(r)\frac{d}{dr}R_{l m \omega }(r)\right]\nonumber\\
&=& \lim_{r\to \infty} \frac{\Delta(r)}{8\pi} \sum_{lm} (-im\Omega) \left[ {\mathfrak R}_{lm\omega}(r)\frac{d}{dr}{\mathfrak R}^*_{lm\omega}(r) -  {\mathfrak R}^*_{lm\omega}(r)\frac{d}{dr}{\mathfrak R}_{lm\omega}(r)\right]_{\omega=m\Omega}\nonumber\\
&=& \lim_{r\to \infty} \frac{\Delta(r)}{4\pi} \sum_{lm} m\Omega\,\,  {\mathcal  Im}\left({\mathfrak R}_{lm\omega}(r)\frac{d}{dr}{\mathfrak R}^*_{lm\omega}(r)\right)\big|_{\omega=m\Omega}\,.
\eea
\end{widetext}
Let us introduce the constants (i.e., not depending on $r$)
\bea
{\mathfrak F }_{lm}(r_0)&=& \lim_{r\to \infty} \Delta(r) {\mathcal  Im}\left({\mathfrak R}_{lm\omega}(r)\frac{d}{dr}{\mathfrak R}^*_{lm\omega}(r)\right)\big|_{\omega=m\Omega}\nonumber\\
F^{\rm up}_{lm}(r_0)&=& \lim_{r\to \infty} \Delta(r) {\mathcal  Im}\left(R_{\rm up}(r)\frac{d}{dr}R^*_{\rm up}\right)\big|_{\omega=m\Omega}\,,
\eea
and the combination
\beq
{\mathcal F }_{lm}(r_0)=
q_{_{\rm TS}}^2 |C_{lm\omega}(r_0)|^2\big|_{\omega=m\Omega}F^{\rm up}_{lm}(r_0)\,.
\eeq
Note that one cannot use the PN solutions for $R_{\rm in/up}(r)$ because these solutions are real, implying necessarily ${\mathcal F }_{lm}(r_0)=0$. In other words,
${\mathcal F }_{lm}(r_0)$ requires the MST solutions to be computed, and hence depends crucially from having imposed the correct boundary conditions for the solution of the radial equation.
 
Consequently,
\bea
\frac{dE}{dt}&=& \frac{q_{_{\rm TS}}^2}{4\pi}\sum_{lm} m\Omega |C_{lm\omega}(r_0)|^2\big|_{\omega=m\Omega}F^{\rm up}_{lm}(r_0)\,,
\eea
with
\bea
 |C_{lm\omega}(r_0)|^2 &=& \frac{16\pi^2} {\Gamma^2}\frac{|R_{\rm in}(r_0)|^2  |Y_{lm}(\frac{\pi}{2},0)|^2}{|W_{lm\omega}|^2}\nonumber\\
&=&  16\pi^2  \left(1-\frac{3r_s}{2r_0}\eta_v^2\right) \times \nonumber\\
&&\times \frac{|R_{\rm in}(r_0)|^2  |Y_{lm}(\frac{\pi}{2},0)|^2}{|W_{lm\omega}|^2}\,.
\eea

In the present investigation we limit our considerations to the $l=0,1,2$ modes,   
\bea
\frac{dE}{dt}&=&\left(\frac{dE}{dt}\right)^{l=0} +\left(\frac{dE}{dt}\right)^{l=1}+\left(\frac{dE}{dt}\right)^{l=2}\,,
\eea
where, for example,
\beq
\left(\frac{dE}{dt}\right)^{l=0}=\frac{q_{_{\rm TS}}^2}{4\pi} \sqrt{\frac{r_s}{2 r_0^{3}}}\lim_{m\to 0}\left(  m |C_{0m\omega}(r_0)|^2\big|_{\omega=m\Omega}F^{\rm up}_{0m}(r_0)\right)\,.
\eeq
Using the notation 
\beq
u=\frac{r_s}{2r_0}\,,\qquad r_b=\alpha r_s\,,
\eeq
and limiting our considerations at the NNLO approximation level we find (omitting the overall factor $q^2/(4\pi^2 R_y^2 M^2)=q_{_{\rm TS}}^2/M^2$)
\bea
\left(\frac{dE}{dt}\right)^{l=0}&=& O(u^8)\,,\nonumber\\
\left(\frac{dE}{dt}\right)^{l=1}&=& -\frac{u^4}{3}\left(1-\frac{2}{5}u (5 \alpha +13) \eta_v ^2+\right.\nonumber\\
&+&\left. \frac{1}{175} u^2(35 \alpha  (5 \alpha +39)+1123) \eta_v ^4+\dots\right)\nonumber\\
&+& O(u^8)\,,\nonumber\\
\left(\frac{dE}{dt}\right)^{l=2}&=& -\frac{16u^5}{15}\left(1-\frac{1}{7}u (28 \alpha +53) \eta_v ^2 +\right.\nonumber\\
&+&\left.\frac{4}{441}u^2 (21 \alpha  (28 \alpha +137)+1906) \eta_v ^4
   +\dots\right)\nonumber\\
&+& O(u^8)\,.
\eea
The sum of the various contributions yields
\bea
    \frac{dE}{dt}
&=& - \left(\frac{q_{_{\rm TS}}}{M}\right)^2 \frac{u^4}{3} \left[ 1-2u-\frac{3117}{175}u^2+\frac{121984}{2205}u^3 \right. \nonumber\\
&+& \alpha\left(-2 u-5u^2+\frac{8768}{105}u^3 \right)\nonumber\\
&+&\left.\alpha^2  \left(u^2+\frac{256}{15}u^3 \right)\right]+O(u^8)\,,
\eea
where $q_{_{\rm TS}} =\frac{q}{2\pi R_y}$,
and  
we put  $\eta_v=1$. Moreover, we have restored physical dimensions, denoting $r_s=2M$, a convenient length scale deduced from the Schwarzschild case (and used to rescale both $E$ and $t$ so that their ratio is dimensionless). Note that, differently from the 4D case where $q\sim L$ has the dimensions of a length, in the present 5D case $q\sim L^2$, {\it i.e.} it scales with $r_s^2$ and not with $r_s$.
In a more compact notation
\bea
\frac{dE}{dt} &=& - \left(\frac{q_{_{\rm TS}}}{M}\right)^2 \frac{u^4}{3} \left[{\mathcal E}_0(u)+\alpha{\mathcal E}_1(u)+
\alpha^2 {\mathcal E}_2(u)\right]\,,\qquad
\eea
where
\bea
{\mathcal E}_0(u)&=& 1-2u-\frac{3117}{175}u^2+\frac{121984}{2205}u^3 +O(u^4)\,,  \nonumber\\
\nonumber\\
{\mathcal E}_1(u)&=& -2 u-5u^2+\frac{8768}{105}u^3 +O(u^4)\,, \nonumber\\
\nonumber\\
{\mathcal E}_2(u)&=& u^2+\frac{256}{15}u^3 +O(u^4)\,,\nonumber\\
\eea
with ${\mathcal E}_k(u)$ starting with $u^k$, $k=0,1,2$.

\begin{figure}
\includegraphics[scale=0.7]{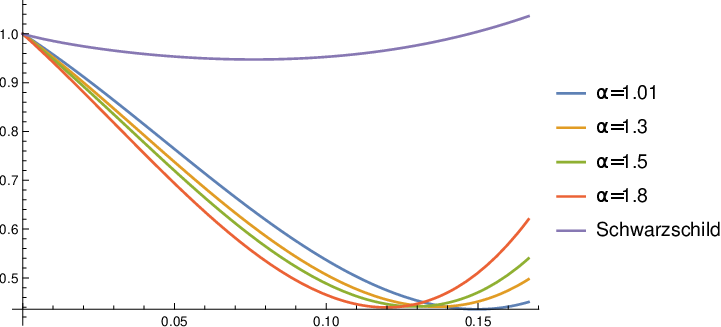}
\caption{\label{fig:energy} Comparison between the radiated energy in Schwarschild (per unit of $q_s/M$) and the TS (per unit of $q_{_{\rm TS}}/M$) as functions of $u$ and for different values of $\alpha$, both rescaled by the factor $(-u^4/3)$. The plots start at $u=0$ with the same values but different inclinations (i.e., $-2$ for Schwarzschild, $-2-2\alpha$ for the TS) and then they differ as soon as both $u$ and $\alpha$ increase. Note that the rescaling produces a minimum which disappears when restoring the factor $(-u^4/3)$.}
\end{figure}

Note that in the Schwarzschild case Ref. \cite{Bini:2016egn}, the first of Eqs. Eq. (5.14) with $y=u$ (the variable  $y$ denotes here the extra dimension and cannot be used), has been found to be
\bea
\frac{dE_{\rm Schw}}{dt}&=&-\left(\frac{q_s}{M}\right)^2\frac13 u^4 \left(1-2u+2\pi u^{3/2}\right.\nonumber\\
&-& \left.10u^2+\frac{12}{5}\pi u^{5/2}\right)+O(u^3)\,.
\eea
Here $q_s$ denotes the scalar charge of the perturbing particle in the Schwarzschild spacetime, different from $q$ since $\frac{q_s}{M}$ is dimensionless.
The difference with the TS case  is actually summarized by the presence of an overall factor $\left(\frac{q_{_{\rm TS}}}{M}\right)^2$ plus $\alpha$-corrections terms.
The $\alpha^0$ LO and NLO terms coincide exactly, but then numerical differences arise. Finally, the fractional power terms are absent at this approximation level and appear at higher PN contributions.

Moreover, because both $t$ and $\phi$ are  coordinates adapted to the Killing vectors of the background, $\xi_{(t)}=\partial_t$ and $ \xi_{(\phi)}=\partial_\phi$, the angular momentum variation along the symmetry axis, $dL$, is simply related to the energy variation $dE$ by (see Eq. (4.13) of Ref. \cite{Teukolsky:1974yv} which also contains the relation between the loss of $L$ and the energy momentum tensor, $dE=T^{\mu\nu} \xi_{(t)\mu}d\Sigma_\nu$, $dL=T^{\mu\nu} \xi_{(\phi)\mu}d\Sigma_\nu$, etc.)
\beq
dL=\frac{m}{\omega}dE\,,
\eeq
where $m$ is the azimuthal quantum number.

\section{Discussion}

Topological Stars  are extremely simple instances of smooth horizonless geometries.
In the present investigation we have studied the energy radiated by a
massive scalar particle orbiting around a TS. Thanks to spherical symmetry the orbit can be taken to lie in the equatorial plane $\theta =\pi/2$ with $\phi=\Omega t$ and $r=r_0$ before including energy losses. Due to emission of massless scalar waves, the massive scalar particle (`probe') experiences self-force effects that make it spiral in (the self force is proportional to $q^2$ and hence independent on the sign of the scalar charge $q$). In order to simplify our analysis we have focussed on the $n=0$ Kaluza-Klein sector that correspond to `smearing' the point-like source along the compact (periodic) $y$ direction ($y \sim y +2\pi R_y$) or to replacing it with a small un-excited string wound around $y$. 

The massless scalar waves produced have been derived using the Green's function method, based on our knowledge of the $up$ and $in$ solutions for the radial motion. 

We then computed the self-force on the massive probe (neutral scalar or string) along its equatorial circular orbit and found that it is purely radial ($F^{\rm orb}_\theta=F^{\rm orb}_y = 0$ in fact also $F^{\rm orb}_\phi=F^{\rm orb}_t = 0$, due to cancellation after summing over $m$) and and constant, independent of $t$ and $\phi$ but only on $r_0$ ($\theta=\pi/2$ is preserved). 
Then we derived the (first-order) corrections to the dynamics giving rise to epicycles.

Taking into account the emission of massless scalar waves we eventually computed the loss in terms of radiated energy and angular momentum, which in this case is simply related to the energy loss. Tortoise coordinate and asymptotics are studied in Appendix A while the visualization of the TS geometry is described in Appendix B.

Finally, we performed a rough comparison with similar results in `standard' BH contexts (Schwarzschild BHs) and commented on extensions to $n\neq 0$ (KK sector) and $s\neq 0$ waves,  e.g.  gravitational. This nontrivial generalization, however, requires further study because it involves a new radial equation with more than four regular singularities, also deviating from Heun type equations \cite{Bena:2024hoh,Dima:2024cok}. 
We hope to report soon on this very interesting problem, partly analyzed from different
perspectives in previous works. 

\section*{Acknowledgments} 
We would like to thank F.~Fucito, A.~Geralico, F.~Morales and A.~Ruiperez-Vicente,  for useful discussions and comments, and in particular M.~Melis and P.~Pani for sharing their preliminary results.  
D.~B. acknowledges sponsorship of the Italian Gruppo Nazionale per la Fisica Matematica
(GNFM) of the Istituto Nazionale di Alta Matematica (INDAM).
M.~B. and G.~D.~R. thank the MIUR PRIN contract 2020KR4KN2 \lq\lq String Theory as a bridge
between Gauge Theories and Quantum Gravity'' and the INFN project ST\&FI \lq\lq String Theory and
Fundamental Interactions'' for partial support.

\appendix

\section{Tortoise coordinates and asymptotic behaviors}

Let us introduce the tortoise coordinate
\begin{equation}
\label{tort}
r_*=r-r_b+(r_b-r_s)\log(r-r_b)\,,  
\end{equation}
satisfying
\beq
\frac{dr_*}{dr}=\frac{r-r_s}{r-r_b}=\frac{f_s(r)}{f_b(r)}\,,
\eeq
and thus mapping the interval $r\in[r_b,\infty)$ in $r_*\in(-\infty, \infty)$.
Using $r_*$ allows to reduce the radial equation in its normal form for the (rescaled) radial function  $ \Psi(r_*)=\sqrt{\Delta} R(r)$,  with an effective potential  given by
\bea
Q_r(r_*)&=& \frac{(r-r_b)[\omega^2 r^3-L(r-r_s)-r_b+r_s]}{(r-r_s)^4}\nonumber\\
&=& \frac{f_b(r)}{r^2 f_s(r)^4}\left(\omega^2 r^2-Lf_s(r)-\frac{f_s(r)-f_b(r)}{r} \right)\,.\nonumber\\
\eea
When $r\to \infty $ ($\sqrt{\Delta}\sim r$) we find
\begin{equation}
Q_r(r_*)\sim \omega^2\,,\qquad R(r_*)\sim \frac{e^{\pm i \omega r_*}}{r}\,,
\end{equation}
 whereas when $r\to r_b$
\beq 
Q_r(r_*)\sim 0\,,\qquad R(r_*)\sim {\rm constant}\,, 
\eeq
with $c_{1,2}$ integration constants.
We can then define the `up' solution at infinity 
\beq
R_{\rm up}^{r\to \infty}=C_{\rm trans}\frac{e^{i\omega r_*}}{r}
\eeq
and the `in' solution at infinity
\beq
R_{\rm in}^{r\to \infty}=B_{\rm ref}\frac{e^{i\omega r_*}}{r}+B_{\rm inc}\frac{e^{-i\omega r_*}}{r}\,.
\eeq
We find
\beq
F^{\rm up}_{lm}(r_0)=-i |C_{\rm trans}|^2\omega\,,\qquad W_{lm\omega}=2i\omega C_{\rm trans} B_{\rm inc}\,,
\eeq
which can be used for a check of the previous quantitative results, after deriving the expressions for $C_{\rm trans}$ and  $B_{\rm inc}$ by using the MST general solutions.

\section{TS spacetime: visualization of the $r-\phi$ geometry via embedding diagrams}
 
Let us consider the $\{r,\phi\}$ section of the metric \eqref{metric_top_star}, namely the metric induced on the equatorial hypersurfaces $t=$constant, $y=$constant (with $\theta=\frac{\pi}{2}$),
\bea
\label{metric_top_star_induc}
ds_{(r,\phi)}^2 &=& \frac{dr^2}{f_s(r)f_b(r)}+r^2  d\phi^2 \nonumber\\
&=&  h_{rr} dr^2+ r^2 d\phi^2  \,.
\eea
An Euclidean embedding diagram for the metric \eqref{metric_top_star_induc} consists in writing it in the form
\bea
ds^2_{\rm E_3}&=& dr^2+r^2 d\phi^2 +dZ(r)^2\nonumber\\
&=& \left[1+\left(\frac{dZ}{dr}\right)^2\right]dr^2 + r^2 d\phi^2\,,
\eea
whereas a Lorentzian (Minkowskian) embedding diagram corresponds to
\bea
ds^2_{\rm M_3}&=& dr^2+r^2 d\phi^2 -dZ(r)^2\nonumber\\
&=& \left[1-\left(\frac{dZ}{dr}\right)^2\right]dr^2 + r^2 d\phi^2\,,
\eea
Both cases are summarized by to a first order differential equation for
$Z(r)$  
\begin{equation}
 1 \pm (dZ(r)/dr)^2 =h_{rr} \ ,
\end{equation}
that is 
\beq\label{eq:emb1}
	|dZ(r)/dr| =\sqrt{ \pm(h_{rr}-1)} \ ,
\eeq
implying
\beq
\label{eq:emb2}
	|Z(r)-Z_0|=\int_{r_0}^r \sqrt{ \pm(h_{rr}-1)} \,dr \ ,
\eeq
where the $\pm$ sign corresponds to the embedding in $E_3$/$M_3$ \cite{Bini:1997ea,Bini:1997eb}.

Once the curve in the  $r$-$Z$ plane is obtained (either analytically
or numerically), the full surface is obtained by revolving it around
the $Z$ axis.  Moreover, when the tangent to the embedding cross-section is not vertical, one
can introduce the tangent cone obtained by revolving the tangent line
at a point $(r,Z(r))$ on the cross-section.

For the TS case the Euclidean embedding can be integrated analytically leading to
\bea
Z(r)&=& 2\sqrt{\frac{f_s(r)}{f_b(r)}}
\sqrt{r \left(r_b+r_s\right)-r_b r_s}+ 2  r_b{\mathcal E}(r)\,,
\eea
(up to an additive arbitrary constant) where
\beq
{\mathcal E}(r)=i{\rm EllipticE}\left(i \sinh ^{-1}\left(\frac{r_b}{\sqrt{r{-}r_b}
   \sqrt{r_b+r_s}}\right)\bigg|1-\frac{r_s^2}{r_b^2}\right)\,,
\eeq
which is real and for example when $r\to \infty$ behaves as
\beq
{\mathcal E}\approx -\frac{r_b}{\sqrt{r(r_b+r_s)}}\,.
\eeq

Figure \ref{fig:3} shows the behavior of this diagram with superposed the one corresponding to the Schwarzschild spacetime.

\begin{figure}
\includegraphics[scale=0.7]{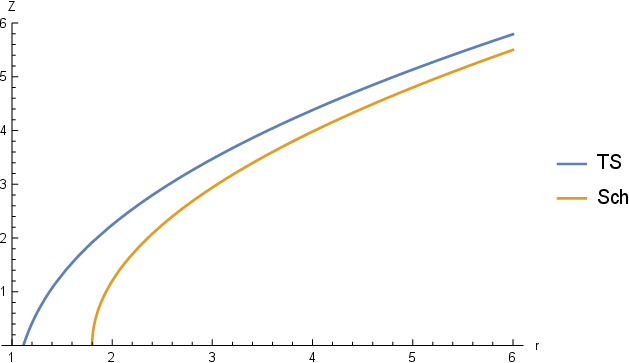}
\caption{\label{fig:3} Plot of the $r-\phi$ section of a  TS spacetime for $r_b=1$, $r_s=0.8$  (blue on-line) superposed the one corresponding to the Schwarzschild spacetime (yellow on-line) for a new $r_s=1.8$ so that they match at large $r$. The additive constant has been fixed here so that both graphs start at $Z=0$ (TS at $r=r_b$ and Schwarzschild at $r=r_s$).}
\end{figure}

It is worth to note that L. Smarr \cite{Smarr:1973zz} has considered embedding diagrams for which the signature changes from Euclidean to Minkowskian  studying the 2-surface cross-section of the
horizon of a charged Kerr black hole. One can repeat the same type of analysis here, extending the TS embedding to the region $r_s <r<r_b$ (Minkowskian embedding) and $0<r<r_s$ (Euclidean embedding) but this would have only a geometrical meaning.

Finally, the Gaussian curvature (represented below as the only nonvanishing component of the Riemann tensor in its mixed form) of this 2D metric is given by
\beq
R^{r\phi}{}_{r\phi}=-\frac{ (r_b+r_s) r-2 r_s r_b }{ 2r^4}\,,
\eeq
with a quadratic scalar invariant
\beq
{\mathcal K}_{(r,\phi)}=R_{\alpha\beta\gamma\delta}R^{\alpha\beta\gamma\delta}=\frac{[(r_b+r_s) r-2 r_s r_b ]^2}{r^8}\,.
\eeq

\begin{figure}
\vskip 0.1 cm
\includegraphics[scale=0.35]{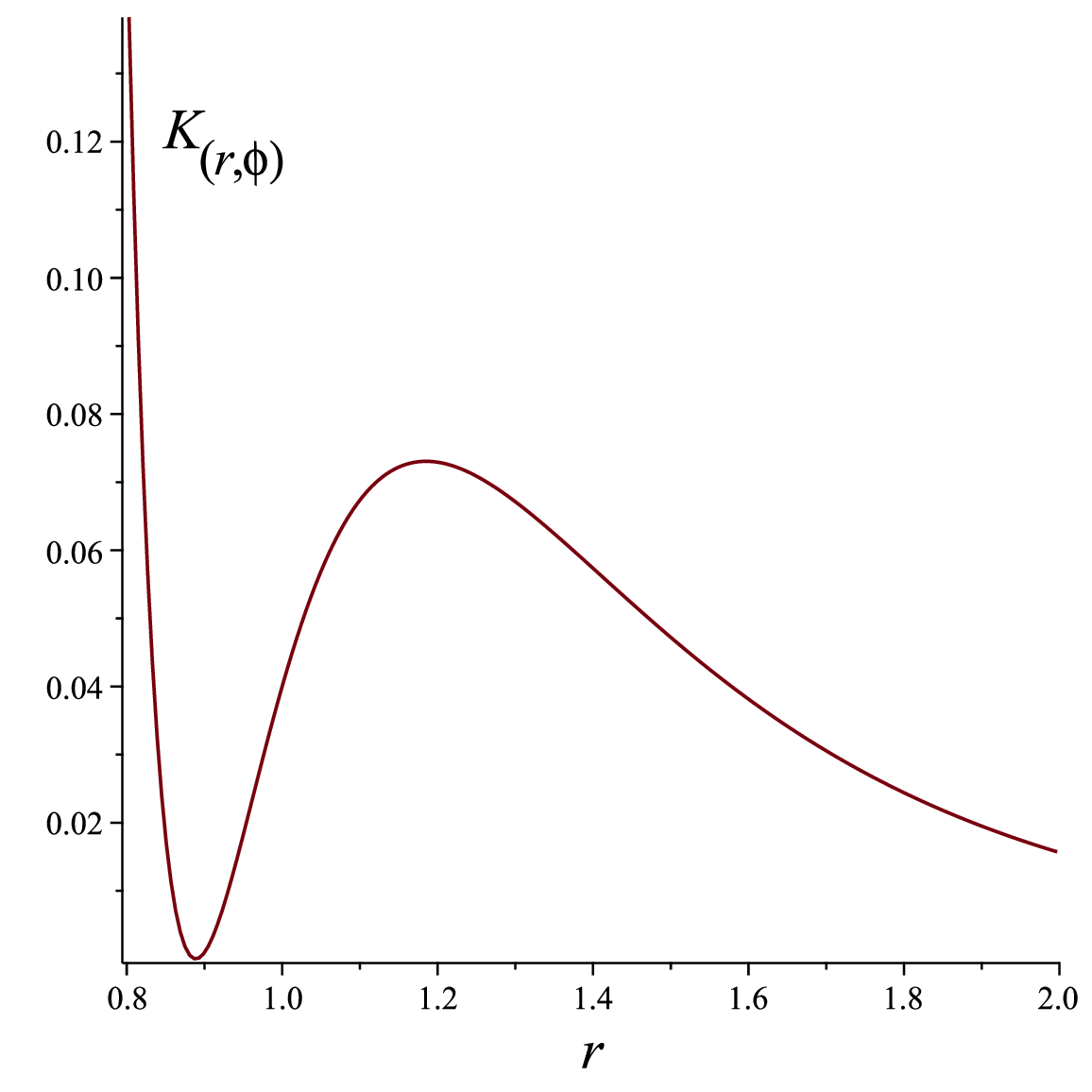}
\caption{Plot of the quadratic scalar invariant ${\mathcal K}_{(r,\phi)}(r)$ as a function of $r$ for $r_b=1$ and $r_s=0.8$. The curve has a minimum at $r=r_{\rm flat}$ (with value ${\mathcal K}_{(r,\phi)}(r_{\rm flat})=0$) and a maximum at $r_{\rm max}=\frac43 r_{\rm flat}$ (with value ${\mathcal K}_{(r,\phi)}(r_{\rm max})=\frac{729 (r_b + r_s)^8 }{ 4194304 r_s^6 r_b^6 }\approx 0.0731$).
}
\end{figure}

The curvature of the $r-\phi$ section is then singular at $r=0$ (outside the region of physical interest, $r>r_b$) and vanishes at the radius $r_{\rm flat}$ (within the region of physical interest for $\alpha>1$) given by
\beq
r_{\rm flat}=\frac{2 r_s r_b}{(r_b+r_s)}=\frac{2\alpha r_s}{1+\alpha}=\frac{1}{\frac12 \left(\frac{1}{r_b}+\frac{1}{r_s} \right)}\,.
\eeq
Evidently, the Ricci scalar is
\bea
R_{(2)}&=&-2R^{r\phi}{}_{r\phi}
\eea
also vanishes at $r=r_{\rm flat}$, featuring the local flatness of the $r-\phi$ section at this radius.

Let us consider now the section $r-y$ with metric (induced on the equatorial hypersurfaces $t=$constant, $\phi=$constant (with $\theta=\frac{\pi}{2}$))
\bea
\label{metric_ry}
ds^2 &=&  \frac{dr^2}{f_s(r)f_b(r)}   + f_b(r) dy^2\nonumber\\
&=& h_{RR}dR^2+R^2dy^2\,,
\eea
with $R^2=f_b(r)$ (dimensionless) that is
\beq
r=\frac{r_b}{1-R^2}
\eeq
and
\bea
h_{RR}(R)&=& \frac{1}{f_s(r)f_b(r)}\left(\frac{dr}{dR}\right)^2\nonumber\\
&=& \frac{4 r_b^3}{(r_b-r_s+r_s R^2) (R^2-1)^4}
\eea
The Euclidean embedding in this case implies
\beq
ds^2 
= \left[1+\left(\frac{dZ}{dR}\right)^2\right]dR^2+R^2dy^2=h_{RR}dR^2+R^2dy^2\,,
\eeq
that is
\beq
\frac{dZ}{dR}=\sqrt{h_{RR}-1}\,,
\eeq
leading to
\beq
Z(R)=\int \sqrt{h_{RR}-1} dR +{\rm const}.
\eeq

The quadratic scalar invariant
\beq
{\mathcal K}_{(r,y)}=R_{\alpha\beta\gamma\delta}R^{\alpha\beta\gamma\delta}=\frac{(4 r-5 r_s)^2r_b^2}{4r^8}\,,
\eeq
and has a minimum (with value 0) at $r_{\rm flat}^{(r,y)}=\frac54 r_s$ and a maximum (with value $\frac{729r_b^2}{62500 r_s^6} $) at $r_{\rm max}^{(r,y)}=\frac53 r_s$.

\begin{figure}
\includegraphics[scale=0.35]{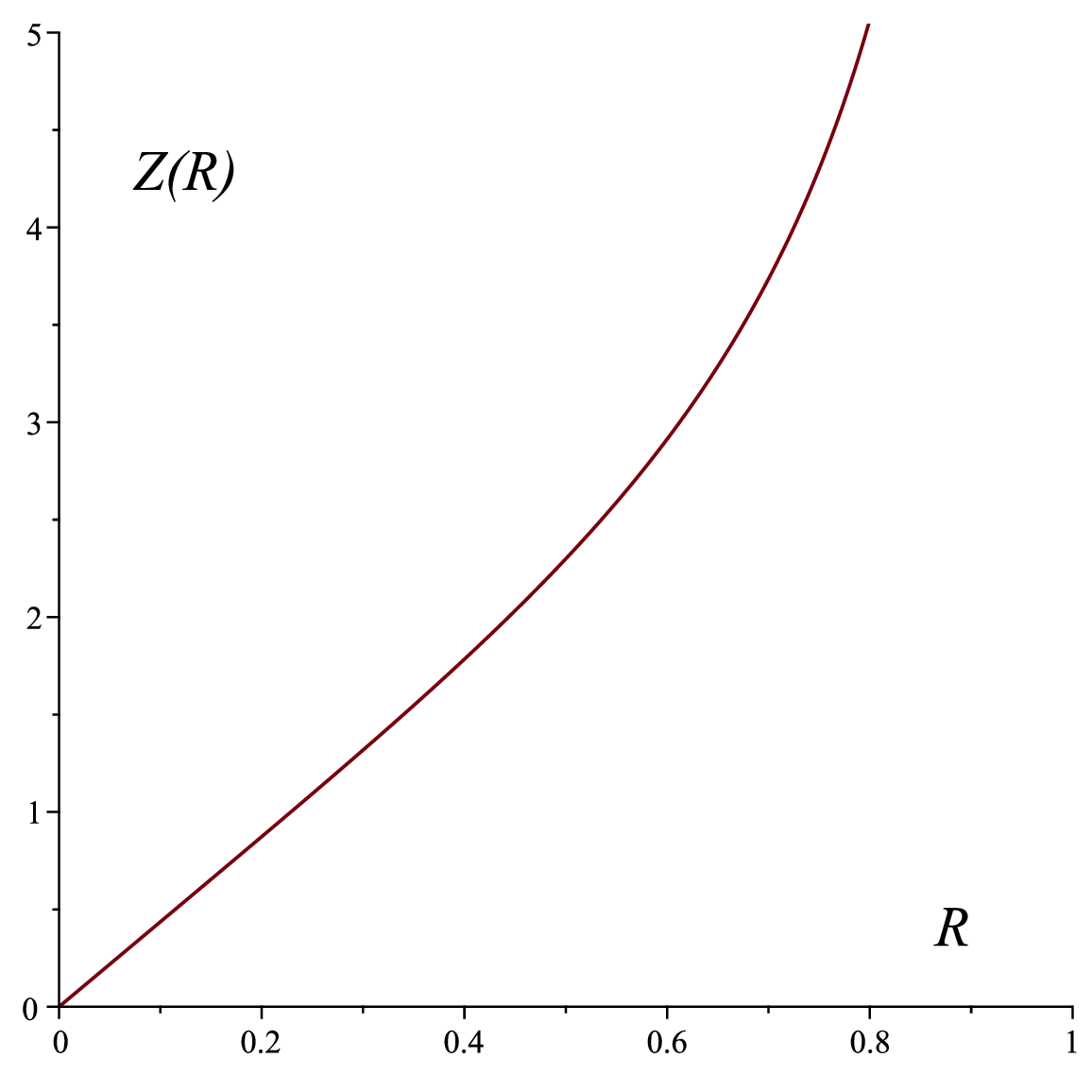}
\caption{\label{fig:3} The Euclidean embedding diagram of the $r-y$ section of a  TS spacetime, for $r_b=1$, $r_s=0.8$.}
\end{figure}


\begin{thebibliography}{99}
 
\bibitem{LigoVirgoKagra}
See the Scientific Collaboration Webpages:
https://ligo.org;
https://www.virgo-gw.eu;
https://gwcenter.icrr.u-tokyo.ac.jp.


\bibitem{Raposo:2018xkf}
G.~Raposo, P.~Pani and R.~Emparan,
``Exotic compact objects with soft hair,''
Phys. Rev. D \textbf{99}, no.10, 104050 (2019)
doi:10.1103/PhysRevD.99.104050
[arXiv:1812.07615 [gr-qc]].

\bibitem{Barack:2018yly}
L.~Barack, V.~Cardoso, S.~Nissanke, T.~P.~Sotiriou, A.~Askar,
C.~Belczynski, G.~Bertone, E.~Bon, D.~Blas and R.~Brito, \textit{et al.}
``Black holes, gravitational waves and fundamental physics: a
roadmap,''
Class. Quant. Grav. \textbf{36}, no.14, 143001 (2019)
doi:10.1088/1361-6382/ab0587
[arXiv:1806.05195 [gr-qc]].


\bibitem{Lunin:2001jy}
O.~Lunin and S.~D.~Mathur,
``AdS / CFT duality and the black hole information paradox,''
Nucl. Phys. B \textbf{623}, 342-394 (2002)
[arXiv:hep-th/0109154 [hep-th]].


\bibitem{Bena:2007kg}
I.~Bena and N.~P.~Warner,
``Black holes, black rings and their microstates,''
Lect. Notes Phys. \textbf{755}, 1-92 (2008)
doi:10.1007/978-3-540-79523-0\_1
[arXiv:hep-th/0701216 [hep-th]].

\bibitem{Skenderis:2008qn}
K.~Skenderis and M.~Taylor,
``The fuzzball proposal for black holes,''
Phys. Rept. \textbf{467}, 117-171 (2008)
doi:10.1016/j.physrep.2008.08.001
[arXiv:0804.0552 [hep-th]].


\bibitem{Bena:2015bea}
I.~Bena, S.~Giusto, R.~Russo, M.~Shigemori and N.~P.~Warner,
``Habemus Superstratum! A constructive proof of the existence of superstrata,''
JHEP \textbf{05}, 110 (2015)
doi:10.1007/JHEP05(2015)110
[arXiv:1503.01463 [hep-th]].

\bibitem{Bianchi:2020miz}
M.~Bianchi, D.~Consoli, A.~Grillo, J.~F.~Morales, P.~Pani and G.~Raposo,
``The multipolar structure of fuzzballs,''
JHEP \textbf{01}, 003 (2021)
[arXiv:2008.01445 [hep-th]].

\bibitem{Bianchi:2020bxa}
M.~Bianchi, D.~Consoli, A.~Grillo, J.~F.~Morales, P.~Pani and G.~Raposo,
``Distinguishing fuzzballs from black holes through their multipolar structure,''
Phys. Rev. Lett. \textbf{125}, no.22, 221601 (2020)
[arXiv:2007.01743 [hep-th]].


\bibitem{Bena:2020uup}
I.~Bena and D.~R.~Mayerson,
``Black Holes Lessons from Multipole Ratios,''
JHEP \textbf{03}, 114 (2021)
[arXiv:2007.09152 [hep-th]].

\bibitem{Bena:2020see}
I.~Bena and D.~R.~Mayerson,
``Multipole Ratios: A New Window into Black Holes,''
Phys. Rev. Lett. \textbf{125}, no.22, 221602 (2020)
[arXiv:2006.10750 [hep-th]].


\bibitem{Bianchi:2017sds}
M.~Bianchi, D.~Consoli and J.~F.~Morales,
``Probing Fuzzballs with Particles, Waves and Strings,''
JHEP \textbf{06}, 157 (2018)
doi:10.1007/JHEP06(2018)157
[arXiv:1711.10287 [hep-th]].

\bibitem{Bianchi:2022qph}
M.~Bianchi and G.~Di Russo,
``2-charge circular fuzz-balls and their perturbations,''
JHEP \textbf{08}, 217 (2023)
doi:10.1007/JHEP08(2023)217
[arXiv:2212.07504 [hep-th]].

\bibitem{DiRusso:2024hmd}
G.~Di Russo, F.~Fucito and J.~F.~Morales,
``Tidal resonances for fuzzballs,''
JHEP \textbf{04}, 149 (2024)
[arXiv:2402.06621 [hep-th]].

\bibitem{Bianchi:2020yzr}
M.~Bianchi, D.~Consoli, A.~Grillo and J.~F.~Morales,
``Light rings of five-dimensional geometries,''
JHEP \textbf{03}, 210 (2021)
doi:10.1007/JHEP03(2021)210
[arXiv:2011.04344 [hep-th]].


\bibitem{Bianchi:2023sfs}
M.~Bianchi, G.~Di Russo, A.~Grillo, J.~F.~Morales and G.~Sudano,
``On the stability and deformability of top stars,''
JHEP \textbf{12}, 121 (2023)
[arXiv:2305.15105 [gr-qc]].


\bibitem{Bianchi:2018kzy}
M.~Bianchi, D.~Consoli, A.~Grillo and J.~F.~Morales,
``The dark side of fuzzball geometries,''
JHEP \textbf{05}, 126 (2019)
doi:10.1007/JHEP05(2019)126
[arXiv:1811.02397 [hep-th]].



\bibitem{Bena:2019azk}
I.~Bena, P.~Heidmann, R.~Monten and N.~P.~Warner,
``Thermal Decay without Information Loss in Horizonless Microstate Geometries,''
SciPost Phys. \textbf{7}, no.5, 063 (2019)
doi:10.21468/SciPostPhys.7.5.063
[arXiv:1905.05194 [hep-th]].

\bibitem{Maldacena:2015waa}
J.~Maldacena, S.~H.~Shenker and D.~Stanford,
``A bound on chaos,''
JHEP \textbf{08}, 106 (2016)
doi:10.1007/JHEP08(2016)106
[arXiv:1503.01409 [hep-th]].

\bibitem{Bianchi:2020des}
M.~Bianchi, A.~Grillo and J.~F.~Morales,
``Chaos at the rim of black hole and fuzzball shadows,''
JHEP \textbf{05}, 078 (2020)
doi:10.1007/JHEP05(2020)078
[arXiv:2002.05574 [hep-th]].


\bibitem{Susskind:2014rva}
L.~Susskind,
``Computational Complexity and Black Hole Horizons,''
Fortsch. Phys. \textbf{64}, 24-43 (2016)
doi:10.1002/prop.201500092
[arXiv:1403.5695 [hep-th]].


\bibitem{Cotler:2016fpe}
J.~S.~Cotler, G.~Gur-Ari, M.~Hanada, J.~Polchinski, P.~Saad,
S.~H.~Shenker, D.~Stanford, A.~Streicher and M.~Tezuka,
``Black Holes and Random Matrices,''
JHEP \textbf{05}, 118 (2017)
[erratum: JHEP \textbf{09}, 002 (2018)]
doi:10.1007/JHEP05(2017)118
[arXiv:1611.04650 [hep-th]].

\bibitem{Bah:2020pdz}
I.~Bah and P.~Heidmann,
``Topological stars, black holes and generalized charged Weyl solutions,''
JHEP \textbf{09}, 147 (2021)
[arXiv:2012.13407 [hep-th]].

\bibitem{Heidmann:2023ojf}
P.~Heidmann, N.~Speeney, E.~Berti and I.~Bah,
``Cavity effect in the quasinormal mode spectrum of topological stars,''
Phys. Rev. D \textbf{108}, no.2, 024021 (2023)
[arXiv:2305.14412 [gr-qc]].

\bibitem{Bianchi:2024vmi}
M.~Bianchi, D.~Bini and G.~Di Russo,
``Scalar perturbations of topological-star spacetimes,''
Phys. Rev. D \textbf{110}, no.8, 084077 (2024)
doi:10.1103/PhysRevD.110.084077
[arXiv:2407.10868 [gr-qc]].

\bibitem{Mano:1996vt}
S.~Mano, H.~Suzuki and E.~Takasugi,
``Analytic solutions of the Teukolsky equation and their low frequency expansions,''
Prog. Theor. Phys. \textbf{95}, 1079-1096 (1996)
[arXiv:gr-qc/9603020 [gr-qc]].


\bibitem{Nekrasov:2009rc}
N.~A.~Nekrasov and S.~L.~Shatashvili,
``Quantization of Integrable Systems and Four Dimensional Gauge Theories,''
[arXiv:0908.4052 [hep-th]].

\bibitem{Nekrasov:2002qd}
N.~A.~Nekrasov,
``Seiberg-Witten prepotential from instanton counting,''
Adv. Theor. Math. Phys. \textbf{7}, no.5, 831-864 (2003)
[arXiv:hep-th/0206161 [hep-th]].

 
\bibitem{Alday:2009aq}
L.~F.~Alday, D.~Gaiotto and Y.~Tachikawa,
``Liouville Correlation Functions from Four-dimensional Gauge Theories,''
Lett. Math. Phys. \textbf{91}, 167-197 (2010)
[arXiv:0906.3219 [hep-th]].


\bibitem{Bianchi:2021xpr}
M.~Bianchi, D.~Consoli, A.~Grillo and J.~F.~Morales,
``QNMs of branes, BHs and fuzzballs from quantum SW geometries,''
Phys. Lett. B \textbf{824}, 136837 (2022)
[arXiv:2105.04245 [hep-th]].


\bibitem{Bianchi:2021mft}
M.~Bianchi, D.~Consoli, A.~Grillo and J.~F.~Morales,
``More on the SW-QNM correspondence,''
JHEP \textbf{01}, 024 (2022)
[arXiv:2109.09804 [hep-th]].



\bibitem{Consoli:2022eey}
D.~Consoli, F.~Fucito, J.~F.~Morales and R.~Poghossian,
``CFT description of BH\textquoteright{}s and ECO\textquoteright{}s: QNMs, superradiance, echoes and tidal responses,''
JHEP \textbf{12}, 115 (2022)
[arXiv:2206.09437 [hep-th]].

\bibitem{Fucito:2023afe}
F.~Fucito and J.~F.~Morales,
``Post Newtonian emission of gravitational waves from binary systems: a gauge theory perspective,''
JHEP \textbf{03}, 106 (2024)
[arXiv:2311.14637 [gr-qc]].


\bibitem{Cipriani:2024ygw}
A.~Cipriani, C.~Di Benedetto, G.~Di Russo, A.~Grillo and G.~Sudano,
``Charge (in)stability and superradiance of Topological Stars,''
[arXiv:2405.06566 [hep-th]].

\bibitem{Aminov:2020yma}
G.~Aminov, A.~Grassi and Y.~Hatsuda,
``Black Hole Quasinormal Modes and Seiberg\textendash{}Witten Theory,''
Annales Henri Poincare \textbf{23}, no.6, 1951-1977 (2022)
[arXiv:2006.06111 [hep-th]].

\bibitem{Aminov:2023jve}
G.~Aminov, P.~Arnaudo, G.~Bonelli, A.~Grassi and A.~Tanzini,
``Black hole perturbation theory and multiple polylogarithms,''
JHEP \textbf{11}, 059 (2023)
[arXiv:2307.10141 [hep-th]].


\bibitem{Bianchi:2023rlt}
M.~Bianchi, C.~Di Benedetto, G.~Di Russo and G.~Sudano,
``Charge instability of JMaRT geometries,''
JHEP \textbf{09}, 078 (2023)
[arXiv:2305.00865 [hep-th]].


\bibitem{Bautista:2023sdf}
Y.~F.~Bautista, G.~Bonelli, C.~Iossa, A.~Tanzini and Z.~Zhou,
``Black hole perturbation theory meets CFT2: Kerr-Compton amplitudes from Nekrasov-Shatashvili functions,''
Phys. Rev. D \textbf{109}, no.8, 084071 (2024)
[arXiv:2312.05965 [hep-th]].

\bibitem{Bonelli:2022ten}
G.~Bonelli, C.~Iossa, D.~Panea Lichtig and A.~Tanzini,
``Irregular Liouville Correlators and Connection Formulae for Heun Functions,''
Commun. Math. Phys. \textbf{397}, no.2, 635-727 (2023)
[arXiv:2201.04491 [hep-th]].

\bibitem{Bonelli:2021uvf}
G.~Bonelli, C.~Iossa, D.~P.~Lichtig and A.~Tanzini,
``Exact solution of Kerr black hole perturbations via CFT2 and instanton counting: Greybody factor, quasinormal modes, and Love numbers,''
Phys. Rev. D \textbf{105}, no.4, 044047 (2022)
[arXiv:2105.04483 [hep-th]].


\bibitem{Bena:2024hoh}
I.~Bena, G.~Di Russo, J.~F.~Morales and A.~Ruip\'erez,
``Non-spinning tops are stable,''
[arXiv:2406.19330 [hep-th]].

\bibitem{Dima:2024cok}
A.~Dima, M.~Melis and P.~Pani,
Phys. Rev. D \textbf{110}, no.8, 084067 (2024)
doi:10.1103/PhysRevD.110.084067
[arXiv:2406.19327 [gr-qc]].


\bibitem{Detweiler:2002gi}
S.~L.~Detweiler, E.~Messaritaki and B.~F.~Whiting,
``Selfforce of a scalar field for circular orbits about a Schwarzschild black hole,''
Phys. Rev. D \textbf{67}, 104016 (2003)
doi:10.1103/PhysRevD.67.104016
[arXiv:gr-qc/0205079 [gr-qc]].

\bibitem{Nakano:2003he}
H.~Nakano, N.~Sago and M.~Sasaki,
``Gauge problem in the gravitational selfforce. 2. First postNewtonian force under Regge-Wheeler gauge,''
Phys. Rev. D \textbf{68}, 124003 (2003)
doi:10.1103/PhysRevD.68.124003
[arXiv:gr-qc/0308027 [gr-qc]].


\bibitem{Bini:2016egn}
D.~Bini, G.~Carvalho and A.~Geralico,
``Scalar field self-force effects on a particle orbiting a Reissner-Nordstr\"om black hole,''
Phys. Rev. D \textbf{94}, no.12, 124028 (2016)
doi:10.1103/PhysRevD.94.124028
[arXiv:1610.02235 [gr-qc]].

\bibitem{Teukolsky:1974yv}
S.~A.~Teukolsky and W.~H.~Press,
``Perturbations of a rotating black hole. III - Interaction of the hole with gravitational and electromagnet ic radiation,''
Astrophys. J. \textbf{193}, 443-461 (1974)
doi:10.1086/153180

\bibitem{Bini:1997ea}
D.~Bini, P.~Carini and R.~T.~Jantzen,
``The Intrinsic derivative and centrifugal forces in general relativity. 1. Theoretical foundations,''
Int. J. Mod. Phys. D \textbf{6}, 1-38 (1997)
doi:10.1142/S0218271897000029
[arXiv:gr-qc/0106013 [gr-qc]].

\bibitem{Bini:1997eb}
D.~Bini, P.~Carini and R.~T.~Jantzen,
``The Intrinsic derivative and centrifugal forces in general relativity. 2. Applications to circular orbits in some familiar stationary axisymmetric space-times,''
Int. J. Mod. Phys. D \textbf{6}, 143-198 (1997)
doi:10.1142/S021827189700011X
[arXiv:gr-qc/0106014 [gr-qc]].

\bibitem{Smarr:1973zz}
L.~Smarr,
``Surface Geometry of Charged Rotating Black Holes,''
Phys. Rev. D \textbf{7}, 289-295 (1973)
doi:10.1103/PhysRevD.7.289




  


\end{thebibliography}
\end{document}